\documentclass[3p,twocolumn]{elsarticle}

\journal{Control Engineering Practice}

\usepackage{lineno,hyperref}
\modulolinenumbers[5]

\usepackage{graphicx} % for pdf, bitmapped graphics files
\usepackage{epsfig} % for postscript graphics files
\usepackage{amsmath} % assumes amsmath package installed
\usepackage{amssymb}  % assumes amsmath package installed
\usepackage{epstopdf}
\usepackage{mathtools}
\usepackage{multirow}
\usepackage{array}
\usepackage{siunitx}
\usepackage{tikz}
\usepackage{mathrsfs}
\usepackage{tabularx}
\usepackage{amsthm}
\usepackage{url}
\usetikzlibrary{calc}
\usetikzlibrary{positioning}
\usetikzlibrary{arrows}
\usetikzlibrary{shapes}
\usetikzlibrary{fit}
\newtheorem{rem}{Remark}
\newtheorem{assm}{Assumption}

\newtheorem{theorem}{Theorem}

\usepackage{algpseudocode}
\usepackage{float}
\newcommand{\pv}[1]{\hat #1}
\newcommand{\ltvpv}[1]{\hat #1}

\newcommand{\boost}{\textrm{boost}}
\newcommand{\EGR}{\textrm{EGR}}
\newcommand{\steady}[1]{\bar #1}
\newcolumntype{L}[1]{>{\raggedright\let\newline\\\arraybackslash\hspace{0pt}}m{#1}}
\newcolumntype{C}[1]{>{\centering\let\newline\\\arraybackslash\hspace{0pt}}m{#1}}
\newcolumntype{R}[1]{>{\raggedleft\let\newline\\\arraybackslash\hspace{0pt}}m{#1}}
\usepackage{booktabs}

\usetikzlibrary{shapes.misc}

\tikzset{cross/.style={cross out, draw=black, minimum size=2*(#1-\pgflinewidth), inner sep=0pt, outer sep=0pt},
	cross/.default={1pt}}

\DeclareMathOperator{\diag}{diag}

\begin{document}
	
		%% Normalising constants %%

		\def\pimbar{200}
	\def\egrbar{100}
	\def\NOxbar{600}
	\def\OPbar{100}

		\title{Model Predictive Controller with Average \\ Emissions Constraints for Diesel Airpath \tnoteref{t1}}
		
		\tnotetext[t1]{Funding: This work was supported by Australian Research Council (ARC) [Grant number - LP160100650], Toyota Motor Corporation, Japan and The University of Melbourne through Melbourne International Research Scholarship (MIRS) Melbourne-India Postgraduate Program (MIPP).}

		\author[mech]{Gokul S.~Sankar\corref{cor}}
		\ead{ggowri@student.unimelb.edu.au}
		
		\author[mech]{Rohan C.~Shekhar}
		\ead{rshekhar@unimelb.edu.au}
		
		\author[elec]{Chris Manzie}
		\ead{manziec@unimelb.edu.au}
		
		\author[tmc]{Takeshi Sano}
		\ead{takeshi\_sano\_aa@mail.toyota.co.jp}
		
		\author[tmc]{Hayato Nakada}
		\ead{hayato\_nakada@mail.toyota.co.jp}
		
		\cortext[cor]{Corresponding author}
		\address[mech]{Department of Mechanical Engineering, The University of Melbourne, Victoria 3010, Australia.}
		\address[elec]{Department of Electrical and Electronic Engineering, The University of Melbourne, Victoria 3010, Australia.}
		\address[tmc]{Advanced Unit Management System Development Division, Toyota Motor Corporation, Higashi-Fuji Technical Center, 1200, Mishuku, Susono-city, Shizuoka, 410-1193 Japan.}

	%============================================================================================%

	\begin{abstract}
		
		Diesel airpath controllers are required to deliver good tracking performance whilst satisfying operational constraints and physical limitations of the actuators. Due to explicit constraint handling capabilities, model predictive controllers (MPC) have been successfully deployed in diesel airpath applications. Previous MPC implementations have considered instantaneous constraints on engine-out emissions in order to meet legislated emissions regulations. However, the emissions standards are specified over a drive cycle, and hence, can be satisfied on average rather than just instantaneously, potentially allowing the controller to exploit the trade-off between emissions and fuel economy. In this work, an MPC is formulated to maximise the fuel efficiency whilst tracking boost pressure and exhaust gas recirculation (EGR) rate references, and in the face of uncertainties, adhering to the input, safety constraints and constraints on emissions averaged over some finite time period. The tracking performance and satisfaction of average emissions constraints using the proposed controller are demonstrated through an experimental study considering the new European drive cycle.
		
	\end{abstract}


	\maketitle

\section{Introduction}

In the development of control systems for diesel airpath applications, it is challenging to achieve a successful trade-off between drivability and fuel efficiency whilst satisfying legislated emission limits. Aftertreatment systems such as diesel particulate filter (DPF) and selective catalytic reduction (SCR) have been introduced in the diesel engine in order to treat engine-out exhaust gas such that the tailpipe Nitrogen oxides $\left(\textrm{NO}_{\textrm{x}}\right)$ and particulate matter (PM) emissions adhere to increasingly stringent regulated levels \cite{Schaer2003, Stewart2008}. In order to achieve optimal performance of the aftertreatment systems, engine-out emissions should be limited to certain levels. This requires close tracking of the reference values for the exhaust gas recirculation (EGR) rate which is defined as the ratio of the EGR outflow rate to the combined EGR and compressor outflow rates.

In addition to tracking EGR rate reference, the diesel airpath controller is required to track reference values for boost pressure for ensuring responsiveness to driver demands whilst minimising pumping losses to improve the fuel efficiency and satisfying operating constraints on intake and exhaust manifold pressures, physical limitations of the actuators. The reference values for boost pressure and EGR rate, for a given engine operating condition characterised by an engine rotational speed, $\omega_e$, and a fuelling rate, $\dot{m}_f$, are determined by a high level controller in order to obtain `optimal' driver demand responsiveness and satisfy emission regulations. The actuators in the diesel airpath, namely  the throttle valve, the EGR valve and the variable geometry turbocharger (VGT) manipulate the flows of fresh air and exhaust gas that influence boost pressure and EGR rate.

The multivariable nature and its ability to systematically handle constraints, makes model predictive control (MPC) an ideal choice of control architecture for constrained multi-input multi-output systems, such as diesel engines. Through simulation studies, application of nonlinear MPC to diesel airpath has been shown to have better control performance compared to the traditional control loops \cite{Herceg2006} but still cannot be implemented in standard engine controllers. Typical sampling rates used in production engine control have motivated the application of linear \cite{Rueckert2004} and explicit MPC formulations \cite{Ortner2007,Ferreau2007,Karlsson2010,Huang2013a,Huang2016}. Techniques including intermittent constraint enforcement \cite{Huang2013}, and imposing soft constraints to ensure controller feasibility \cite{Karlsson2010,Huang2016,Wahlstroem2013} to reduce the computational complexity, lead to loss of guarantees on constraint adherence. Furthermore, since these studies did not account for model imperfections, robust constraint satisfaction guarantees are lost. 

Robust MPC formulations based on tube MPC and constraint tightening approach were proposed for diesel airpath application by \cite{Huang2014} and diesel generator in power tracking application by \cite{Broomhead2017}. However, to achieve certain desired output transient response, the previous implementations require significant calibration effort due to the high number of tuning parameters and the non-intuitive relationship between the parameters and the time domain characteristics of the output response, such as overshoot and settling time.

An MPC formulation for diesel airpath application with a suitable cost function parameterisation and an appropriate controller structure proposed by \cite{Sankar2015, Shekhar2017, Sankar2017} has reduced number of effective tuning parameters that helps in reducing the calibration effort. Based on this MPC formulation, \cite{Sankar2018} proposed a robust switched linear time-invariant (LTI) MPC architecture with multiple linear models \cite{Ortner2006, DelRe2010} to handle drive cycle operation. However, \cite{Sankar2018} did not provide satisfaction guarantees of legislated emissions limits.

This paper is a significant extension of \cite{Sankar2018} that did not consider constraints on emissions. In this work, a robust switched LTI-MPC architecture is proposed that incorporates constraints on engine-out emissions. Conservative static maps of engine-out $\textrm{NO}_{\textrm{x}}$ level and opacity are identified as a function of the states and inputs and used to enforce emissions constraints. Maintaining engine-out emissions under certain levels that can be handled by the aftertreatment systems will result in retaining the tailpipe emissions within regulated levels. The regulated emission limits are typically defined over drive cycles \cite{Commission2018}. Therefore, instead of pointwise-in-time emissions constraints, emissions averaged over drive cycles are considered in this work. Adopting the methodology introduced by \cite{Mueller2014} to handle transient average constraints in an MPC formulations, upper bound constraints on engine-out emissions averaged over some finite time are imposed in the proposed controller. Furthermore, the controller proposed in this work incorporates a penalty on transient pumping loss in the MPC cost metric, allowing fuel economy to be targeted for improvement within the allowable emission limits. 

\subsection{Notation}
The symbol $\mathbb{R}$ represents a set of real numbers. The symbol $\mathbb{Z}_{\left[a: b\right]}$ denotes a set of consecutive integers from $a$ to $b$ and $2\mathbb{Z}^{+}$ denotes set of positive even integers. $0_{m\times n}$ represents a zero matrix of size $m\times n$, $I_n$ denotes an $n\times n$ identity matrix and $\boldsymbol{1}_n$ is a $n \times 1$ vector of ones. The operator $\det\left(A\right)$ denotes the determinant of the matrix $A$. $A \succ 0$ represents a positive definite matrix $A$. The Euclidean norm of a vector $x$ is denoted by $\|x\|$; $\|x\|_1$ represents its $L_1$ norm; and $\|A\|_{\max}\coloneqq\underset{ij}{\max}\,\left|a_{ij}\right|$, where $\left|a_{ij}\right|$ is the absolute value of the element in $i^{th}$ row and $j^{th}$ column of the matrix $A$. The operator $\ominus$ denotes the Pontryagin difference, defined for sets $\mathcal{A}$ and $\mathcal{B}$ as $\mathcal{A}\ominus\mathcal{B}\coloneqq\left\{ a|a+b\in\mathcal{A}\,\forall b\in\mathcal{B}\right\}$ for which the property,
\begin{equation}
c\in\mathcal{A}\ominus\mathcal{B}\,\Rightarrow c+b\in\mathcal{A}\,\forall b\in\mathcal{B},\label{eq:pontdiff}
\end{equation}
\noindent is satisfied. The operator $\diag\{\cdot\}$ denotes a diagonal matrix with the elements in parentheses along the leading diagonal. All inequalities involving vectors are to be interpreted row-wise.

%============================================================================================%
\section{Diesel airpath and emissions modelling}
\label{sec:modelling}	

\begin{figure}
	\begin{centering}
		\begin{tikzpicture}
		\node at (0,0)
		{\includegraphics[clip, trim = {0.cm 0.0cm 0.cm 0.cm}]{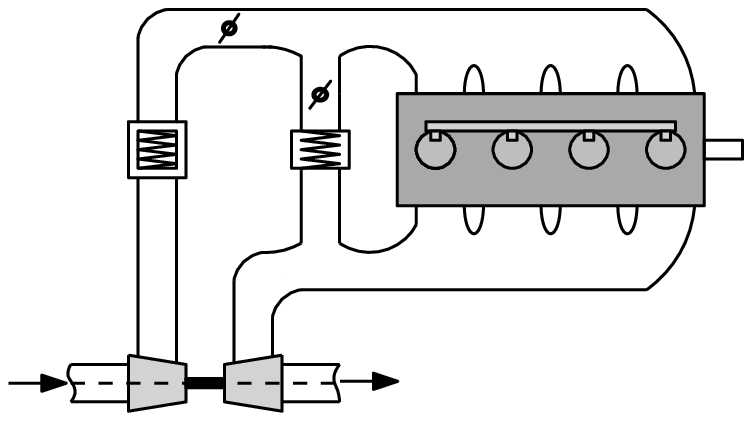}};
		\node at (-3.33,0.65) {Intercooler};
		\node at (-1.33,0.65) {EGR};
		\node at (-1.33,0.375) {cooler};
		\node at (-1.25,1.35) {EGR};
		\node at (-1.25,1.075) {valve};
		\node at (-3.5,-1.4) {Air};
		\node at (-1.5,2.25) {Throttle};
		\node [text width = 2cm, align = center] at (0.6,-1.3)  {Engine-out exhaust gas};
		\node at (1.5,-0.55) {Exhaust manifold};
		\node at (1.5,1.75) {Intake manifold};
		\node at (1.75,0.25) {Cylinders};
		\node at (-2.75,-2.25) {Compressor};
		\node at (-1.2,-2.25) {VGT};
		\node at (1.77,1.025) {\tiny Fuel rail and injectors};
		\end{tikzpicture}
		\par\end{centering}
	\protect\caption{Diesel engine schematic.}
	\label{fig:eng_sch}
\end{figure}

Fig.~\ref{fig:eng_sch} represents a schematic of the airpath of a diesel engine with the positioning of the actuators and other components such as intercooler, cooled EGR system and VGT. The density of the fresh air entering the airpath is first increased by the compressor and then by the intercooler. The high-density air in which more oxygen is available, will help in efficient combustion of the fuel injected from the high pressure rail into the cylinders. 

A portion of the burnt gas in the exhaust manifold flows through the EGR cooler and the EGR valve into the intake manifold. The fresh air-burnt gas mixture has decreased oxygen availability and an increased specific heat capacity which reduces the peak combustion temperature, thereby, reducing NO$_\textrm{x}$ and increasing PM formations. The engine-out exhaust gas drives the VGT, whose shaft spins the compressor. The nozzle geometry at the inlet of the turbine can be varied to influence the flow through the VGT.

\begin{figure}
	\begin{centering}
		\begin{tikzpicture} [scale = .9, transform shape]
		\node at (0,0) (pic) {\includegraphics[clip, trim = {0.65cm 0.55cm 0cm 0.0cm}]{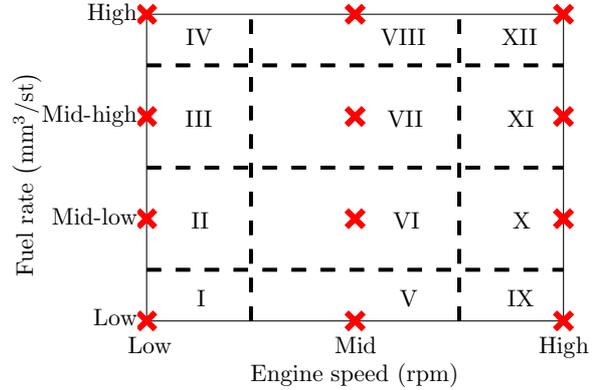}};
		\node [above left = -50mm and -16mm of pic]{I};
		\node [above left = -38.5mm and -16.5mm of pic]{II};
		\node [above left = -23.5mm and -17mm of pic]{III};
		\node [above left = -11.5mm and -17mm of pic]{IV};
		
		\node [above left = -50mm and -47mm of pic]{V};
		\node [above left = -38.5mm and -47.3mm of pic]{VI};
		\node [above left = -23.5mm and -47.8mm of pic]{VII};
		\node [above left = -11.5mm and -48.2mm of pic]{VIII};
		
		\node [above left = -50mm and -64mm of pic]{IX};
		\node [above left = -38.5mm and -63.5mm of pic]{X};
		\node [above left = -23.5mm and -64mm of pic]{XI};
		\node [above left = -11.5mm and -64.5mm of pic]{XII};
		
		%% Labelling Y axis %%
		
		\node (rect) [draw =none, fill = white, minimum width = 0.65cm, minimum height = 4.95cm, inner sep = 0pt, below  left = -54mm and -4.75mm of pic] {};
		
		\node [ below  left = -9mm and -6mm of pic] {Low};
		\node [ below  left = -24mm and -6mm of pic] {Mid-low};
		\node [ below  left = -39mm and -6mm of pic] {Mid-high};
		\node [ below  left = -54mm and -6mm of pic] {High};

		%% Labeling the time axis %%
		
		\node (rect) [draw = none, fill = white, minimum width = 6.75cm, minimum height = 0.3cm, inner sep = 0pt, below  left = -4mm and -71mm of pic] {};
		
		\node [ below  left = -5mm and -11mm of pic] {Low};
		\node [ below  left = -5mm and -41mm of pic] {Mid};
		\node [ below  left = -5mm and -72mm of pic] {High};
		
		\node [above left = -13mm and 8mm of pic, rotate = 90]{Fuel rate $\left(\SI{}{mm^3/st}\right)$};
		\node [below left = -1mm and -53mm of pic, rotate = 0]{Engine speed $\left(\SI{}{rpm}\right)$};
		
		\end{tikzpicture}
		\par\end{centering}
	\caption{Engine operational space divisions and the corresponding linearisation points.}
	\label{fig:mod_grid}
\end{figure}

The diesel airpath is a highly nonlinear system and hence, a single linear model to approximate the behaviour over the entire engine operating range will have low fidelity. Therefore, the engine operating range is divided into $12$ regions as shown in Fig.~\ref{fig:mod_grid} and a fourth-order linear perturbation model about a selected operating point in each region is used to represent the dynamics \cite{Sankar2018}. These linearisation/model grid points are evenly spaced with a resolution of $\SI{800}{rpm}$ and $\SI{25}{mm^3/st}$ on the engine speed and fuelling rate, respectively. A fourth-order model is used as it was found to provide a good balance between computational complexity and fidelity. Furthermore, complete state feedback is available through estimating the EGR rate in the engine control unit (ECU) and directly measuring the other states.

For a given model grid point, $\left(\omega_e^g,\dot{m}_f^g\right)$, the steady state inputs are chosen as the actuator commands applied by the ECU at that operating condition, where $g\in \left\{\text{I},\,\text{II},\,\ldots,\,\text{XII} \right\}$ represents the model grid point. These steady inputs are then applied on the engine to obtain the trimming conditions for the linear models. The steady values for the input, state and output are given by the vectors ${\steady{u}}^g \in \mathbb{R}^3$, ${\steady{x}}^g \in \mathbb{R}^4$ and ${\steady{y}}^g \in \mathbb{R}^2$, respectively.

The linear perturbation model at a given model grid point, $\left(\omega_e^g,\dot{m}_f^g\right)$, that describes the deviations from trim conditions is represented by
\begin{subequations}
	\begin{eqnarray}
	x_{k+1} & = & A_{}^gx_{k}+B^g_{}u_{k}+w_{k},\\
	y_{k} & = & C_{}^gx_{k} + D_{}^gu_{k},
	\end{eqnarray}\label{eq:lsys}
\end{subequations}

\noindent where the perturbed states, $x_{k} \coloneqq \left\{p_{\text{im}}\: p_{\text{em}}\: W_{\textrm{comp}}\: y_{\text{EGR}}\right\}^T$ are respectively the perturbations in the intake and the exhaust manifold pressures, the flow rate through compressor and the EGR rate about ${\steady{x}}^g$; $y_{k}\coloneqq\left\{p_{\text{im}}\: y_{\text{EGR}}\right\}^T$ are the output perturbations; the perturbed control inputs, $u_{k}\coloneqq \left\{u_{\text{thr}}\: u_{\text{EGR}}\: u_{\text{VGT}} \right\}^T$ are respectively the perturbations in the throttle and the EGR valve positions and the VGT position about ${\steady{u}}^g$; and $w_k\in\mathbb{R}^{4}$ is an unknown but bounded state disturbance contained in $\mathcal{W}^g$.

The model parameters are identified by using the engine test bench data as in \cite{Sankar2018} and \textsc{Matlab}'s system identification toolbox. The state disturbance set ${\mathcal{W}^g} \coloneqq  \left\{ w|\zeta^g w\leq\theta^g,\,\zeta^g\in\mathbb{R}^{a\times 4},\,\theta^g\in\mathbb{R}^{a}\right\}$, with $a \in 2\mathbb{Z}^{+}$, is chosen as the hypercube that captures the discrepancies between the linear model predictions and the engine behvaiour arising due to external disturbances, measurement errors and modelling errors as a consequence of using low-order discretised models. The disturbance set for each model grid point is estimated from the test bench data obtained for system identification. These disturbance sets are compact and include the origin.

\begin{assm} Each pair $\left(A^g,\,B^g\right)$ is stabilisable $\forall g\in \left\{\text{I},\,\text{II},\,\ldots,\,\text{XII} \right\}$.			\label{ass:stab}
\end{assm}

Linear static maps of NO$_\textrm{x}$ $\left(\si{ppm}\right)$ and opacity $\left(\%\right)$ are approximated from the experimental data as a function of perturbed states and inputs of the airpath model in \eqref{eq:lsys}:
\begin{align}
	v_{k}  =  C_{v}x_{k} + D_{v}u_{k},		
	\end{align}	\label{eq:lsys_em}
	
	\noindent	where $v_{k} \coloneqq \left\{\textrm{NO}_\textrm{x}\: \textrm{OP}\right\}^T$ denotes the perturbations in NO$_\textrm{x}$ emission and opacity about the steady state emissions represented by $\steady{v}^g$. In this work, opacity (OP) of the engine-out exhaust gases is used as a substitute for PM measurements because of the relatively low cost of OP measurements \cite{Majewski2013} and the correlation between OP and PM. 
	
\begin{figure}
			\begin{centering}
				\begin{tikzpicture}
				\node at (0,0) (pic)
				{\includegraphics[clip, trim = {0.0cm 0cm 0cm 0cm}]{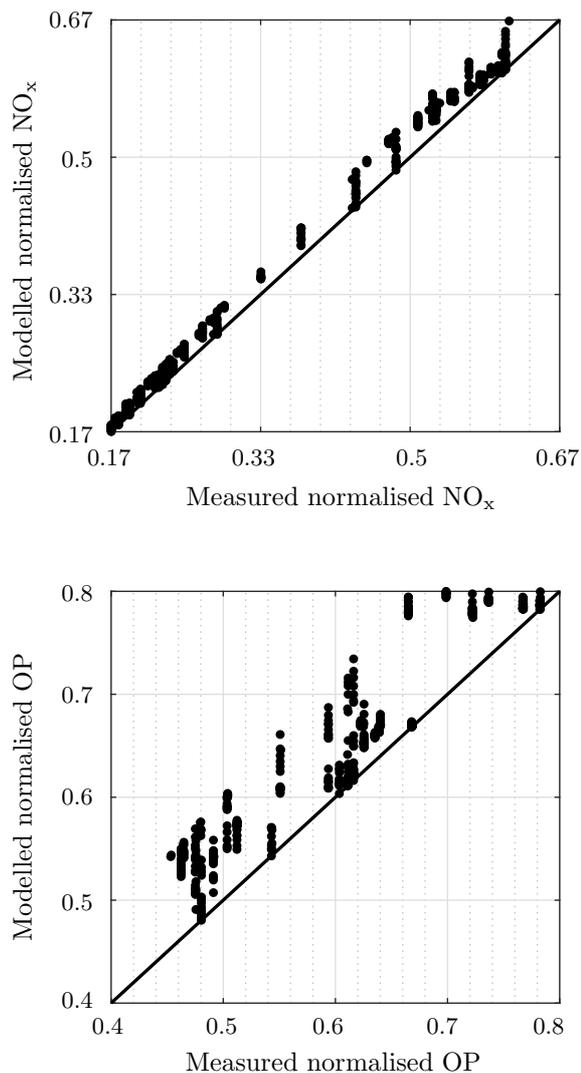}};

				%% Normalising the y axis %%
				
				\node (rect) [ draw = none, fill = white, minimum width = 1.2cm, minimum height = 13.4cm, inner sep = 0pt, below  left = -143mm and -13mm of pic] {};
				
				\node [ below  left = -143mm and -13mm of pic] {\small{\pgfmathparse{400/\NOxbar}\pgfmathprintnumber\pgfmathresult}};
				\node [ below  left = -124.5mm and -13mm of pic] {\small{\pgfmathparse{300/\NOxbar}\pgfmathprintnumber\pgfmathresult}};
				\node [ below  left = -106.5mm and -13mm of pic] {\small{\pgfmathparse{200/\NOxbar}\pgfmathprintnumber\pgfmathresult}};
				\node [ below  left = -88mm and -13mm of pic] {\small{\pgfmathparse{100/\NOxbar}\pgfmathprintnumber\pgfmathresult}};
				
				\node [ rotate = 90, below left = -134mm and 0.0mm of pic] {Modelled normalised $\textrm{NO}_\textrm{x}$};
				
				\node [ below  left = -67mm and -13mm of pic] {\small{\pgfmathparse{80/\OPbar}\pgfmathprintnumber\pgfmathresult}};
				\node [ below  left = -53.5mm and -13mm of pic] {\small{\pgfmathparse{70/\OPbar}\pgfmathprintnumber\pgfmathresult}};
				\node [ below  left = -40mm and -13mm of pic] {\small{\pgfmathparse{60/\OPbar}\pgfmathprintnumber\pgfmathresult}};
				\node [ below  left = -26mm and -13mm of pic] {\small{\pgfmathparse{50/\OPbar}\pgfmathprintnumber\pgfmathresult}};
				\node [ below  left = -13mm and -13mm of pic] {\small{\pgfmathparse{40/\OPbar}\pgfmathprintnumber\pgfmathresult}};
				
				\node [ rotate = 90, below left = -58mm and 0mm of pic] {Modelled normalised OP};
				
							%% Normalising the x axis %%
				
				\node (rect) [draw = none , fill = white, minimum width = 6.7cm, minimum height = 0.9cm, inner sep = 0pt, below  left = -84mm and -76mm of pic] {};
				
				\node (rect) [draw = none , fill = white, minimum width = 6.7cm, minimum height = 0.9cm, inner sep = 0pt, below  left = -9mm and -76mm of pic] {};

				\node [ below  left = -85mm and -18mm of pic] {\small{\pgfmathparse{100/\NOxbar}\pgfmathprintnumber\pgfmathresult}};
				\node [ below  left = -85mm and -37mm of pic] {\small{\pgfmathparse{200/\NOxbar}\pgfmathprintnumber\pgfmathresult}};
				\node [ below  left = -85mm and -57mm of pic] {\small{\pgfmathparse{300/\NOxbar}\pgfmathprintnumber\pgfmathresult}};
				\node [ below  left = -85mm and -77mm of pic] {\small{\pgfmathparse{400/\NOxbar}\pgfmathprintnumber\pgfmathresult}};

				\node [ below left  = -80mm and -66mm of pic] {Measured normalised $\textrm{NO}_\textrm{x}$};

				\node [ below  left = -9.5mm and -17mm of pic] {\small{\pgfmathparse{40/\OPbar}\pgfmathprintnumber\pgfmathresult}};
				\node [ below  left = -9.5mm and -32mm of pic] {\small{\pgfmathparse{50/\OPbar}\pgfmathprintnumber\pgfmathresult}};
				\node [ below  left = -9.5mm and -46.0mm of pic] {\small{\pgfmathparse{60/\OPbar}\pgfmathprintnumber\pgfmathresult}};
				\node [ below  left = -9.5mm and -61.mm of pic] {\small{\pgfmathparse{70/\OPbar}\pgfmathprintnumber\pgfmathresult}};
				\node [ below  left = -9.5mm and -75.0mm of pic] {\small{\pgfmathparse{80/\OPbar}\pgfmathprintnumber\pgfmathresult}};

				\node [ below left  = -5mm and -64mm of pic] {Measured normalised OP};
				
				\end{tikzpicture}
				\par\end{centering}
			\protect\caption{Correlations between measured data and model predictions for $\textrm{NO}_\textrm{x}$ (top) and $\textrm{OP}$ (bottom).}
			
			\label{fig:em_model_corr}
		\end{figure}

	A conservative linear map is identified in this work, which ensures that the predictions from the linear map are greater than the experimental data points by solving:
	\begin{subequations}
		\begin{align}
		\underset{\theta_v}{\min}\,\,\, & \sum_{k=1}^{N_v}\left\Vert \tilde{v}_k - v_k \right\Vert^{2}  \\
		\text{s.t } & \forall k\in\mathbb{Z}_{\left[1:N_v\right]} \nonumber\\
		& {v}_{k}  =  C_{v}\left(\theta_v\right){x}_{k} + D_{v}\left(\theta_v\right){u}_{k} \\
		& \tilde{v}_k - v_k \leq 0,
		\end{align}
		\label{eq:lsq}
	\end{subequations}

	\noindent where $N_v$ is the number of data points and $\tilde{v}_k$ represents the perturbations of the measured NO$_\textrm{x}$ and opacity about $\steady{v}^g$.	The correlations between measured $\textrm{NO}_\textrm{x}$ and $\textrm{OP}$ and the corresponding predictions from the model \eqref{eq:lsys_em} are shown in Fig. \ref{fig:em_model_corr}. It can be noted that the estimates upper bound the emissions data.

%============================================================================================%
\section{Controller development}
\label{sec:controller}

	In this section, a robust model predictive control algorithm will be developed for the diesel airpath in order to regulate the outputs to their reference values, whilst minimising transient pumping loss to improve fuel efficiency and satisfying actuator limitations, safety and reliability constraints and constraints on emissions averaged over some duration. 
	
	The pointwise-in-time state and input constraints to be satisfied at each time instant are given as
\begin{subequations}
	\begin{align}
	&x(k)\in\mathcal{X}\coloneqq \left\{ x|Ex\leq f,\, E\in\mathbb{R}^{q\times 4},\, f\in\mathbb{R}^{q}\right\}, \\
	&u(k)\in\mathcal{U}\coloneqq\left\{ u|Gu\leq h,\, G\in\mathbb{R}^{r\times 3},\, h\in\mathbb{R}^{r}\right\},
	\end{align}\label{eq:con}
\end{subequations}

\noindent where $x(k) = x_k + \steady{x}^g$;  $u(k) = u_k + \steady{u}^g$; $q$ and $r$ represent the number of facets of $\mathcal{X}$ and $\mathcal{U}$, respectively. 

\begin{assm} 
	There exists a positively invariant set, $\mathcal{X}_{f} \coloneqq  \left\{ x|S x\leq t,\,S\in\mathbb{R}^{p\times 4},\,t\in\mathbb{R}^{p}\right\} \subseteq \mathcal{X}$, under a stabilising controller $\kappa_f\left(x\right)$ $\forall \left(A^g,\, B^g\right) $ and $g\in \left\{\text{I},\,\text{II},\,\ldots,\,\text{XII} \right\}$ \cite{Broomhead2014}; where the number of facets of $\mathcal{X}_{f}$ is denoted by $p$.
	\label{ass:terminal_set}
\end{assm}

Let the upper bound on NO$_\textrm{x}$ and opacity averaged over certain finite time period $T\geq1$ be
\begin{align}
\sum_{k=0}^{T} \frac{v(k)}{T} \in \mathcal{V}\coloneqq \left\{ v|\Lambda v\leq \lambda,\, \Lambda \in\mathbb{R}^{2\times 2},\, \lambda \in\mathbb{R}^{2}\right\},
\label{eq:avgcon}
\end{align}
\noindent where $v(k) = v_k + \steady{v}^g$. While considering the average emissions over a drive cycle, $T$ will be equal to the time span of the drive cycle.

The MPC formulation incorporates constraint envelopes on the outputs to reduce the number of effective tuning parameters to assist with rapid calibration \cite{Sankar2018}. This controller structure is used as the basis for designing a controller to address the objectives of this work. The height of the envelope about $\steady{y}^g$ is denoted by $Y_{k+j|k} \in \mathbb{R}^2$ and the  decay  envelope is given by
\begin{equation}
Y_{k+1+j|k}=\Gamma^g Y_{k+j|k},\,\forall j\in\mathbb{Z}_{\left[0:N-2\right]},
\end{equation}
\noindent where $\Gamma^g \coloneqq \exp \left(-\diag \{T_s/\tau_{\boost}^g, T_s/\tau_{\EGR}^g \} \right)$ with the sampling time $T_s$. The primary objective of the controller is to minimise the envelope heights. The envelope time constants, $\tau_{\textrm{boost}}^g$ and $\tau_{\textrm{EGR}}^g$ corresponding to the two output channels can be used as the primary tuning parameters to shape the output response. The secondary objective is to encourage smooth output transients. It is achieved by penalising the output deviations from a nominal exponential decay towards the origin with weighting matrix, $\boldsymbol{\epsilon^{g}} \coloneqq \diag\{\epsilon_{\boost}^{g}, \epsilon_{\EGR}^{g}$\}. The tunable parameters $\epsilon_{\boost}^{g}$ and $\epsilon_{\EGR}^{g}$ are used to adjust the smoothness on corresponding output channel.

The pumping loss is defined as
\begin{align}
\pv{p}^{\textrm{loss}} = \frac{p_{\textrm{em}} - p_{\textrm{im}} - \bar{p}^{\textrm{loss}}_{\min}}{w_{\text{boost}}},
\label{eq:ploss}
\end{align}

\noindent where $w_{\text{boost}}$ is a normalisation constant. The term $\bar{p}^{\textrm{loss}}_{\min}$ is chosen sufficiently small such that $\pv{p}^{\textrm{loss}}$ is positive over the engine operational range, thus, while the pumping loss is minimised as the tertiary objective in the MPC formulation to decrease fuel consumption, pumping gains are not penalised. A weighting parameter on the pumping loss, $\alpha$,  is chosen sufficiently small such that it does not dominate the envelope and smoothness costs. Finally, the deviation of the perturbed inputs from the origin (or equivalently, the deviation of the actual inputs from their steady state values) is penalised with the least priority. Therefore, the MPC cost function is defined as 
\begin{align}
V_{N}\left(x(k),\,\boldsymbol{\pv{u}_k}\right)& \coloneqq  \sum_{j=0}^{N-1}\left\Vert W^g\, Y_{k+j|k}\right\Vert^{2}     \nonumber\\
& \hspace{-.5cm}+	\sum_{j=0}^{N-2} \left\Vert\boldsymbol{\epsilon^{g}} W^g\left(\pv{y}_{k+1+j|k}-\Gamma^g \pv{y}_{k+j|k}\right)\right\Vert^{2} \nonumber \\
& \hspace{-.5cm}+	\sum_{j=0}^{N-1} \left[\alpha \left\Vert \pv{p}^{\textrm{loss}}_{k+j|k}\right\Vert^{2} + \gamma \left\Vert\pv{u}_{k+j|k}\right\Vert^2\right],
\label{eq:mpccost}
\end{align}

\noindent where $N$ is the prediction horizon;  $\boldsymbol{\pv{u}_k} \coloneqq \left\{\pv{u}_{k|k},\pv{u}_{k+1|k},\ldots,\pv{u}_{k+N-1|k}\right\}$ is the input sequence; for a model grid point $g$, $W^{g} \coloneqq \diag\left\{w^{g}/w_{\text{boost}},\,\left(1-w^{g}\right)/w_{\text{EGR}}\right\}$ is the weighting matrix for output prioritisation with the envelope priority parameter $w^{g} \in [0,1]$ and normalisation constants, $w_{\textrm{boost}}$ and $w_{\textrm{EGR}}$; $\gamma \in [0,\infty)$ scales the input regularisation term. The relative priority in minimising the envelope height corresponding to one output at the expense of increase in the envelope height of the other output channel is achieved by tuning the parameter $w^{g}$.

The switched LTI-MPC strategy proposed in \cite{Sankar2018} is utilised in this work to handle the transient operation of the engine. Hence, the system matrices, $A^{g}$, $B^{g}$, $C^{g}$ and $D^{g}$; the steady values, ${\steady{x}_k}$, ${\steady{u}_k}$, ${\steady{y}_k}$ and ${\steady{v}_k}$, which are determined using linear interpolation of the steady state values at the neighbouring grid points; the tuning parameters, $\tau_{\boost}^{g}$, $\tau_{\EGR}^{g}$, $w^{g}$, $\epsilon_{\boost}^{g}$ and $\epsilon_{\EGR}^{g}$; and the constraint tightening margins for state and input constraints,  $\boldsymbol{\sigma}^{g*}\coloneqq \left\{ \sigma_{0}^{g*},\,\sigma_{1}^{g*},\,\ldots,\,\sigma_{N}^{g*}\right\}$ and $\boldsymbol{\mu}^{g*}\coloneqq \left\{ \mu_{0}^{g*},\,\mu_{1}^{g*},\,\ldots,\,\mu_{N-1}^{g*}\right\}$, respectively, are updated at each sampling instant while solving the following MPC optimisation problem,
	\begin{subequations}
	\begin{align}
	\mathcal{P}_N&\left(x(k),\,g\right):  \underset{\boldsymbol{\pv{u}_k},\,Y_{k|k}}{\min}\,\,\,  V_{N}\left(x(k),\,\boldsymbol{\pv{u}_k}\right) \\
	\text{s.t } & \forall j\in\mathbb{Z}_{\left[0:N-1\right]} \nonumber\\
	& \pv{x}_{k|k}=x(k) - {\steady{x}_k} \label{eq:initst}\\
	& \pv{x}_{k+1+j|k}=A^g\pv{x}_{k+j|k}+B^g\pv{u}_{k+j|k} \label{eq:dyn1}\\
	& \pv{y}_{k+j|k}=C^g\pv{x}_{k+j|k}+D^g\pv{u}_{k+j|k} \label{eq:dyn2}\\
	& \pv{v}_{k+j|k}  =  C_{v}\pv{x}_{k+j|k} + D_{v}\pv{u}_{k+j|k} \label{eq:emdyn}\\
	& Y_{k+1+j|k}=\Gamma^g Y_{k+j|k} ,\,\forall j\in\mathbb{Z}_{\left[0:N-2\right]} \label{eq:decay}\\
	& Y_{k|k-1}\leq Y_{k|k} \label{eq:decaylimit}\\
	& -Y_{k+j|k}\leq \pv{y}_{k+j|k}\leq Y_{k+j|k} \label{eq:decaycon}\\
	& E\left(\pv{x}_{k+j|k}+\steady{x}_k\right)\leq f-\sigma^{g*}_{j}\label{eq:statecon}\\
	& G\left(\pv{u}_{k+j|k}+\steady{u}_k\right)\leq h- \mu^{g*}_{j}\label{eq:inpcon} \\
	& S\left(\pv{x}_{k+N|k}+\steady{x}_k\right)\leq t \label{eq:termcon}\\
	& \Lambda\left[\sum_{i =0}^{k-1}v\left(i\right)+\sum_{i =0}^{j_v}\left(\pv{v}_{k+i|k} + \steady{v}_k\right)\right] \nonumber \\ 
	& \hspace{0.2cm} \leq \left(k+j_v+1\right) \lambda, \, j_v = \max \left(0,\,\min \left(j,\, T-k\right)\right) \label{eq:avg1}\\
	& \left|\pv{u}_{k|k-1}^* - \pv{u}_{k|k}\right| \leq \delta \boldsymbol{1}_{3} \label{eq:inpslew1}\\
	& \Delta \pv{u}_{k+j|k} \leq \delta \boldsymbol{1}_3,\,\forall j\in\mathbb{Z}_{\left[0:N-2\right]}\label{eq:inpslew2},
	\end{align}
	\label{eq:mpc}
\end{subequations}

%& \Lambda\sum_{i =j}^{b}\left(\pv{v}_{k+i|k} + \steady{v}_k\right) \leq b \lambda, \nonumber \\
%& \hspace{1.5cm} b = \min \left\{j+T-1,\, N-1\right\} \label{eq:avg2}\\

\noindent where the control input applied to the engine is determined from the control law, $\kappa_N\left(x(k)\right) = \pv{u}^*_{k|k} + {\steady{u}_k}$, in which $\pv{u}^*_{k|k}$ is the first term of the optimal input sequence $\boldsymbol{\pv{u}_k}^* \coloneqq\left\{\pv{u}_{k|k}^*,\pv{u}_{k+1|k}^*,\ldots,\,\pv{u}_{k+N-1|k}^*\right\}$; $v\left(i\right) \coloneqq \pv{v}^*_{i|i}+\steady{v}_i,\,\forall i\in \mathbb{Z}_{\left[0:k-1\right]}$ with $\pv{v}^*_{i|i}$ is obtained from the optimal sequence $\boldsymbol{\pv{v}_i}^* \coloneqq\left\{\pv{v}_{i|i}^*,\pv{v}_{i+1|i}^*,\ldots,\,\pv{v}_{i+N-1|i}^*\right\}$; $\Delta \pv{u}_{k+j|k} \coloneqq \left|\pv{u}_{k+1+j|k} - \pv{u}_{k+j|k}\right|$; $\delta \in \mathbb{R}$ defines the maximal allowable change in the actuator position in one sampling time. 

 The initial condition, nominal system dynamics, the emissions map and the envelope dynamics are included in \eqref{eq:initst}-\eqref{eq:decay}. The constraint \eqref{eq:decaylimit} enforces a condition on the initial envelope height between successive steps of the MPC iteration. The maximum decay rate of the envelopes is restricted to $\Gamma^g$ to prevent one of the outputs decaying too abruptly. Envelope constraints on the output perturbations are enforced through \eqref{eq:decaycon}. Constraint tightening is applied on the state and input constraints through \eqref{eq:statecon} and \eqref{eq:inpcon} to obtain $\left\{\mathcal{X}_0^g,\,\mathcal{X}_1^g,\,\ldots,\,\mathcal{X}_{N}^g\right\}$ and $\left\{\mathcal{U}_0^g,\,\mathcal{U}_1^g,\,\ldots,\,\mathcal{U}_{N-1}^g\right\}$, respectively. The reserved margins, $\boldsymbol{\sigma}^{g*}$ and $\boldsymbol{\mu}^{g*}$, provide constraint satisfaction guarantees for the possible disturbances from the maximal disturbance set \cite{Sankar2018}.

In this work, a $N_{np}$-step nilpotent constraint tightening policy is used, where $N_{np}\leq N$. As the effect of the disturbance entering at the beginning of the horizon can be eliminated in $N_{np}$ steps by utilising the nilpotent policy, no tightening is required for the terminal state constraint in \eqref{eq:termcon}. In \eqref{eq:avg1}, the terms in the first sum are the predictions of the emissions obtained by applying the control law, $\kappa_N\left(x(k)\right)$, up to the time instant $k-1$ and the second term sums up the predictions of the emissions over $k+j_v$ steps, where $j_v = \max \left(0,\,\min \left(j,\, T-k\right)\right)$ $\forall j \in\mathbb{Z}_{\left[0:N-1\right]}$. Finally, the slew rate constraints on the inputs are imposed through \eqref{eq:inpslew1}-\eqref{eq:inpslew2}. The slew rate, $\delta$, is chosen such that $\delta \geq \max \left( \|\mu_{j+1}^{g*} - \mu_{j}^{g*}\|_1\right),\,\forall j \in\mathbb{Z}_{\left[0:N-2\right]}$ and $g\in \left\{\textrm{I},\,\textrm{II},\,\ldots,\,\textrm{XII} \right\}$.

\begin{assm} 
The terminal controller, $\kappa_f({x})$, is chosen such that, $\forall {x}\in\mathcal{X}_f$ and $g\in \left\{\text{I},\,\text{II},\,\ldots,\,\text{XII} \right\}$,
	\begin{enumerate}
		\item $\kappa_f({x}) \in \mathcal{U}_{N-1}^{g}$,
		\item  $C_{v}x + D_{v}\kappa_f({x}) + \steady{v}_k \in \mathcal{V}$.
	\end{enumerate} 
\label{ass:terminal_cntrl}
\end{assm}

\begin{rem} Assumption~\ref{ass:terminal_cntrl} ensures that the terminal controller satisfies the tightened input constraint at the end of the horizon and the emissions are not greater than the upper bound on the average emissions in the terminal region under the terminal controller.	
\end{rem}

\begin{theorem}[Recursive feasibility] Consider that Assumptions \ref{ass:stab} - \ref{ass:terminal_cntrl} hold. If $\mathcal{P}_{N}\left(x(k),\,g\right)$ is feasible, then successive optimisation problems $\mathcal{P}_{N}  \left(x(k+j),\,g'\right)$, are feasible $\forall \,j>0$,  where $g'\in \left\{\textrm{I},\,\textrm{II},\,\ldots,\,\textrm{XII} \right\}$ represents a model grid point.	
	\label{thm:feas}
\end{theorem}

\begin{theorem}[Practical stability]
	Consider the system represented by \eqref{eq:lsys}, subjected to the constraints \eqref{eq:con} and \eqref{eq:avgcon}. Let the Assumptions \ref{ass:stab} - \ref{ass:terminal_cntrl} hold and $\mathbb{X}_N$ be the feasible region for $\mathcal{P}_N(x(k),\,g)$. Then given $N_{np}=1$, a constant trim point $\bar{x}_0$, $x(0)\in\mathbb{X}_N$ and the control law $\kappa_N\left(x(k)\right)$, there exists a class $\mathcal{K}\mathcal{L}$ function $\beta(\cdot,\cdot)$ such that $\forall k\geq0$: 
	
		\begin{eqnarray}		
\hspace{-.7cm}	\left| x(k) - \bar{x}_k\right| \leq &\beta\left(\left|x(0) - \bar{x}_0\right|,\, k\right) \qquad \qquad \qquad \nonumber \\
	& + \mathcal{O} \left(\left\Vert\boldsymbol{\epsilon^{g}}\right\Vert^{2} +\left\Vert \alpha \right\Vert^{2}+\left\Vert \kappa_f \right\Vert^{2} \right).
	\label{eq:stab}
	\end{eqnarray}	
	\label{thm:stab}
\end{theorem}	

\begin{rem} Since stability guarantees are provided about a given steady state condition, Theorem \ref{thm:stab} considers constant trim conditions obtained for a certain engine speed and fuelling rate.
\end{rem}

The proofs of Theorems \ref{thm:feas} and \ref{thm:stab} can be found in the Appendix.

%============================================================================================%
\section{Simulation study}
\label{sec:sim}

	In this section, the proposed controller is implemented in simulations on a high fidelity diesel airpath model and the effect of the average constraints on the emissions obtained over urban driving cycle (UDC) is investigated.  The fixed cost function parameters in \eqref{eq:mpccost} are chosen as $\gamma = \SI{5e-3}{}$, $\alpha = 10^{-2}$, $w_{\boost} = \SI{40}{kPa}$ and $w_{\EGR} = 0.6$. The length of the MPC prediction and control horizons are equal and chosen as N=4. The sampling rate used in this work is consistent with that of the production ECUs. The maximal disturbance set and the corresponding constraint-tightening margins for the state and input constraints are obtained for each model grid point \cite{Sankar2018}. The tuning parameters are chosen as:  $\tau_{\text{boost}}^{g} = 0.5$, $\tau_{\text{EGR}}^{g} = 0.5$, $w^{g}= 0.5$, $\epsilon_{\text{boost}}^{g} = 0$ and $\epsilon_{\text{EGR}}^{g} = 0$,  $\forall g\in \left\{\textrm{I},\,\textrm{II},\,\ldots,\,\textrm{XII} \right\}$.
	
	\begin{figure}
		\begin{centering}
			\begin{tikzpicture}
			\node at (0,0) (pic)
			{\includegraphics[clip, trim = {0.6cm 0cm 0cm 0cm}]{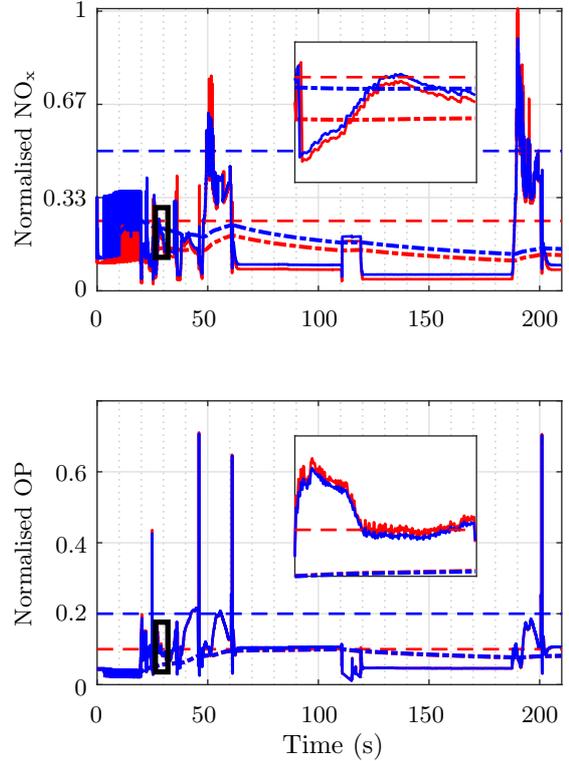}};

				%% Normalising the axis %%
			
			\node (rect) [ draw = none, fill = white, minimum width = 0.9cm, minimum height = 9.5cm, inner sep = 0pt, below  left = -103mm and -7.8mm of pic] {};
			
			\node [ below  left = -103mm and -8mm of pic] {\small{\pgfmathparse{600/\NOxbar}\pgfmathprintnumber\pgfmathresult}};
			\node [ below  left = -91mm and -8mm of pic] {\small{\pgfmathparse{400/\NOxbar}\pgfmathprintnumber\pgfmathresult}};
			\node [ below  left = -78mm and -8mm of pic] {\small{\pgfmathparse{200/\NOxbar}\pgfmathprintnumber\pgfmathresult}};
			\node [ below  left = -66mm and -8mm of pic] {\small{\pgfmathparse{0/\NOxbar}\pgfmathprintnumber\pgfmathresult}};
			
			\node [ rotate = 90, below left = -94mm and 4.0mm of pic] {\small{Normalised {NO$_{\textrm{x}}$}}};
			
			\node [ below  left = -41.5mm and -8mm of pic] {\small{\pgfmathparse{60/\OPbar}\pgfmathprintnumber\pgfmathresult}};
			\node [ below  left = -32mm and -8mm of pic] {\small{\pgfmathparse{40/\OPbar}\pgfmathprintnumber\pgfmathresult}};
			\node [ below  left = -23mm and -8mm of pic] {\small{\pgfmathparse{20/\OPbar}\pgfmathprintnumber\pgfmathresult}};
			\node [ below  left = -13mm and -8mm of pic] {\small{\pgfmathparse{0/\OPbar}\pgfmathprintnumber\pgfmathresult}};
			
			\node [ rotate = 90, below left = -42mm and 4mm of pic] { \small{Normalised {OP}}};
			
			%			%% Labeling the time axis %%
			
			\node (rect) [draw = none , fill = white, minimum width = 6.3cm, minimum height = 0.3cm, inner sep = 0pt, below  left = -61.3mm and -70mm of pic] {};
			
			\node (rect) [draw = none , fill = white, minimum width = 6.3cm, minimum height = 0.8cm, inner sep = 0pt, below  left = -9.3mm and -70mm of pic] {};

			\node [ below  left = -62mm and -10mm of pic] {\small{0}};
			\node [ below  left = -62mm and -25mm of pic] {\small{50}};
			\node [ below  left = -62mm and -41mm of pic] {\small{100}};
			\node [ below  left = -62mm and -55.mm of pic] {\small{150}};
			\node [ below  left = -62mm and -70.mm of pic] {\small{200}};

			\node [ below  left = -9.5mm and -10mm of pic] {\small{0}};
			\node [ below  left = -9.5mm and -25mm of pic] {\small{50}};
			\node [ below  left = -9.5mm and -41.0mm of pic] {\small{100}};
			\node [ below  left = -9.5mm and -55.mm of pic] {\small{150}};
			\node [ below  left = -9.5mm and -70.0mm of pic] {\small{200}};

			\node [ below left  = -6mm and -47mm of pic] {Time (s)};
		
			\end{tikzpicture}
			\par\end{centering}
		\protect\caption{Solid lines denote instantaneous NO$_\textrm{x}$ emissions and opacity; dash-dotted lines represent the cumulative moving average of the emissions; dotted lines denote the upper bound on the emissions averaged over UDC. Blue represents the results obtained for the first case with an upper bound of 0.5 and	0.2, and red is that obtained for the second case with 0.25 and 0.1 on averaged NO$_\textrm{x}$ emissions and opacity, respectively. The inset figures show the magnified views of the corresponding rectangular section.}

		\label{fig:em_udc_sim}
	\end{figure}

				\pgfkeys{/pgf/number format/.cd,fixed,precision=2}
	The instantaneous NO$_\textrm{x}$ emissions and opacity obtained by using a high fidelity emissions model, their corresponding cumulative moving average and the upper bound constraint on the emissions averaged over the drive cycle are shown in Fig~\ref{fig:em_udc_sim} for two choices of upper bounds on the average NO$_\textrm{x}$ emissions and opacity: (i) $\pgfmathparse{300/\NOxbar}\pgfmathprintnumber\pgfmathresult$ and $\pgfmathparse{20/\OPbar}\pgfmathprintnumber\pgfmathresult$; (ii) $\pgfmathparse{150/\NOxbar}\pgfmathprintnumber\pgfmathresult$ and $\pgfmathparse{10/\OPbar}\pgfmathprintnumber\pgfmathresult$, respectively. The upper bound can be adjusted based on the aftertreatment system in-use. 
	
	\begin{figure}
		\begin{centering}
			\begin{tikzpicture}
			\node at (0,0) (pic)
			{\includegraphics[scale = 1, clip, trim = {0.3cm 0cm 0cm 0cm}]{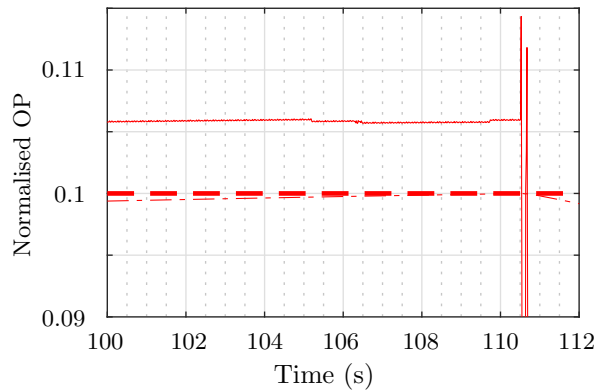}};

			%% Normalising the axis %%
			
			\node (rect) [ draw = none , fill = white, minimum width = .9cm, minimum height = 4.4cm, inner sep = 0pt, below  left = -53.5mm and -10.3mm of pic] {};

			\node [ below  left = -46mm and -9mm of pic] {\small{\pgfmathparse{11/\OPbar}\pgfmathprintnumber\pgfmathresult}};
%			\node [ below  left = -37mm and -8mm of pic] {\small{\pgfmathparse{11/\OPbar}\pgfmathprintnumber\pgfmathresult}};
%			\node [ below  left = -24.5mm and -8mm of pic] {\small{\pgfmathparse{10.5/\OPbar}\pgfmathprintnumber\pgfmathresult}};
			\node [ below  left = -29.5mm and -9mm of pic] {\small{\pgfmathparse{10/\OPbar}\pgfmathprintnumber\pgfmathresult}};
%			\node [ below  left = -24.5mm and -8mm of pic] {\small{\pgfmathparse{9.5/\OPbar}\pgfmathprintnumber\pgfmathresult}};
			\node [ below  left = -13mm and -9mm of pic] {\small{\pgfmathparse{9/\OPbar}\pgfmathprintnumber\pgfmathresult}};
			
			\node [ rotate = 90, below left = -42mm and 3mm of pic] { \small{Normalised {OP}}};
			
			%			%% Labeling the time axis %%
			
			\node (rect) [draw = none , fill = white, minimum width = 7cm, minimum height = 0.8cm, inner sep = 0pt, below  left = -9.3mm and -76mm of pic] {};

			\node [ below  left = -10mm and -14mm of pic] {\small{100}};
			\node [ below  left = -10mm and -25mm of pic] {\small{102}};
			\node [ below  left = -10mm and -35mm of pic] {\small{104}};
			\node [ below  left = -10mm and -45mm of pic] {\small{106}};
			\node [ below  left = -10mm and -55.5mm of pic] {\small{108}};
			\node [ below  left = -10mm and -66mm of pic] {\small{110}};
			\node [ below  left = -10mm and -76mm of pic] {\small{112}};

			\node [ below left  = -6mm and -47mm of pic] {Time (s)};
			
			\end{tikzpicture}
			\par\end{centering}
		\protect\caption{Effect of activity of the averaged constraints on the instantaneous opacity (solid) and its cumulative moving average (dash-dot) trajectories for the second case. Upper bound on the averaged opacity is represented by dotted lines.}
		
		\label{fig:sim_con_act}
	\end{figure}

	In the first case, the peak instantaneous NO$_\textrm{x}$ and opacity of $\pgfmathparse{606.5/\NOxbar}\pgfmathprintnumber\pgfmathresult$ and $\pgfmathparse{70.73/\OPbar}\pgfmathprintnumber\pgfmathresult$ occur at $\SI{190.2}{s}$ and $\SI{46.01}{s}$, respectively. The cumulative moving averages of NO$_\textrm{x}$ and opacity reach a peak value of $\pgfmathparse{152.81/\NOxbar}\pgfmathprintnumber\pgfmathresult$ and $\pgfmathparse{10.01/\OPbar}\pgfmathprintnumber\pgfmathresult$ at $\SI{19.68}{s}$ and $\SI{110.5}{s}$, respectively, while at the end of the drive cycle, the averages are $\pgfmathparse{90.4/\NOxbar}\pgfmathprintnumber\pgfmathresult$ and $\pgfmathparse{8.15/\OPbar}\pgfmathprintnumber\pgfmathresult$, respectively, as reported in Table \ref{tab:avg_sim}. In the other case, where the upper bound on the averaged  NO$_\textrm{x}$ emissions and opacity are reduced to  $\pgfmathparse{150/\NOxbar}\pgfmathprintnumber\pgfmathresult$ and $\pgfmathparse{10/\OPbar}\pgfmathprintnumber\pgfmathresult$, respectively, the cumulative moving average of NO$_\textrm{x}$ reach a peak value of $\pgfmathparse{126.7/\NOxbar}\pgfmathprintnumber\pgfmathresult$ at $\SI{60.78}{s}$. The cumulative moving average of opacity activates the upper bound constraint at $\SI{110.58}{s}$ as shown in Fig.~ \ref{fig:sim_con_act}. It can be noted that the instantaneous emission is greater than $\pgfmathparse{10/\OPbar}\pgfmathprintnumber\pgfmathresult$ before $\SI{110.58}{s}$. However, when the constraint is activated, the controller takes action to reduce the opacity. This deactivates the constraint, allowing room for more opacity and hence, a spike in the opacity can be observed. The peak NO$_\textrm{x}$ and opacity in this case are $\pgfmathparse{605.4/\NOxbar}\pgfmathprintnumber\pgfmathresult$ and $\pgfmathparse{71.09/\OPbar}\pgfmathprintnumber\pgfmathresult$, respectively, with averages of $\pgfmathparse{85.76/\NOxbar}\pgfmathprintnumber\pgfmathresult$ and $\pgfmathparse{8.09/\OPbar}\pgfmathprintnumber\pgfmathresult$, respectively, at the end of the drive cycle. The cumulative moving averages satisfy their respective upper bounds over the drive cycle in both cases.

	\begin{table}
		\pgfkeys{/pgf/number format/.cd,fixed,precision=3}
		\protect\caption{Emissions averaged over UDC and the corresponding pumping losses for different choices of upper bounds on the average emissions.}
		\par
		\centering{}
		\begin{tabular}{>{\centering\arraybackslash} m{0.7cm} >{\centering\arraybackslash} m{1.1cm} >{\centering\arraybackslash} m{0.7cm}  >{\centering\arraybackslash} m{1.1cm}  >{\centering\arraybackslash} m{1.9cm}}
			\toprule
			\multicolumn{2}{ >{\centering\arraybackslash}m{2.2cm}}{Upper bound on average} &  \multicolumn{2}{ >{\centering\arraybackslash}m{2.1cm}}{Average emissions} & \multirow{2}{2.2cm}{Normalised pumping loss} \tabularnewline
			\cline{1-4}
			NO$_\textrm{x}$ & Opacity & NO$_\textrm{x}$ & Opacity & \tabularnewline
			\midrule
			\pgfmathparse{300/\NOxbar}\pgfmathprintnumber\pgfmathresult & \pgfmathparse{20/\OPbar}\pgfmathprintnumber\pgfmathresult & \pgfmathparse{90.4/\NOxbar}\pgfmathprintnumber\pgfmathresult & \pgfmathparse{8.15/\OPbar}\pgfmathprintnumber\pgfmathresult & 1 \tabularnewline
			
			\pgfmathparse{150/\NOxbar}\pgfmathprintnumber\pgfmathresult & \pgfmathparse{10/\OPbar}\pgfmathprintnumber\pgfmathresult & \pgfmathparse{85.76/\NOxbar}\pgfmathprintnumber\pgfmathresult & \pgfmathparse{8.09/\OPbar}\pgfmathprintnumber\pgfmathresult & 1.024 \tabularnewline	
			
			\bottomrule			
			
		\end{tabular}\label{tab:avg_sim}
	\end{table}

	The insets in Fig. \ref{fig:em_udc_sim} show the emissions over a selected time period of the drive cycle. The reduced instantaneous NO$_\textrm{x}$ emissions in the second case (shown in red) is achieved by increased EGR rate that results in more particulate matter emissions indicated by marginally increased opacity and reduced in the fuel economy. In order to compare the fuel economy between the two cases, the pumping losses incurred over the drive cycle are used. As seen from Table~\ref{tab:avg_sim}, reducing the upper bound on averaged emissions in the second case, resulted in an increase of pumping loss of $2.4\%$ compared to the first case. This increased pumping loss can be correlated with an increase in fuel use.
	
	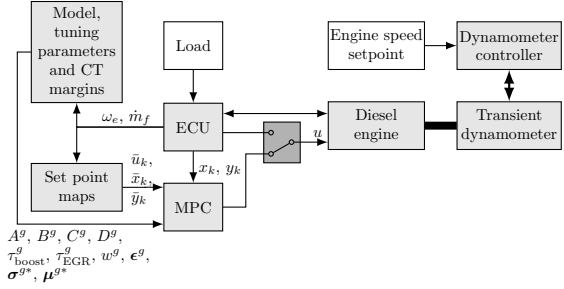
\begin{figure}
		\begin{centering}
			\begin{tikzpicture} [scale = 0.6, transform shape]
			
			\node [text width=6em, minimum height=3em, text centered,draw,fill=gray!20, inner sep = 0pt] (engine) at (0,0) {Diesel \\ engine};
			\node [text width=6.5em, minimum height=3em, text centered,draw,fill=gray!20, right = 7mm of engine, inner sep = 0pt] (dyn)  {Transient dynamometer};
			\node [text width=6.5em, minimum height=3em, text centered,draw,fill=gray!20, above = 7mm of dyn, inner sep = 0pt] (dyncon) {Dynamometer controller};
			\node [text width=6.0em, minimum height=3.0em, text centered,draw, inner sep = 0pt, above = 7mm of engine] (omegaesetpoint) {Engine speed setpoint};
			
			\draw [line width=3pt] (engine) edge (dyn);
			\draw [latex-latex,thick] (dyn) edge (dyncon);
			\draw [-latex] (omegaesetpoint) edge (dyncon);
			
			% switch
			\node [draw, fill=gray!60, below left = -6.20mm and 6mm of engine, minimum width = 0.8cm, minimum height = 0.9cm, inner sep = 0pt] (switch) {};
			\node [draw, circle, minimum size = 1mm, below left = 6.7mm and 2mm of switch.north, inner sep = 0pt] (inportmpc) {};
			\node [draw, circle, minimum size = 1mm, below left = 2.mm and 2mm of switch.north, inner sep = 0pt] (inportecu) {};
			\node [draw, circle, minimum size = 1mm, below left = 4.1mm and -2.5mm of switch.north, inner sep = 0pt] (outport) {};
			\node [text width=4em,text centered, below = 1mm of switch, inner sep = 0pt] (swtichtextbox) {};
			\draw (inportmpc) -- (outport);

			\node [text width=3em, minimum height=3em, text centered,draw,fill=gray!20, above left = -11mm and 23mm of engine] (ecu)  {ECU};
			\node [text width=3.0em, minimum height=3em, text centered,draw, left = 23mm of omegaesetpoint] (load) {Load};
			\node [text width=3em, minimum height=3em, text centered,draw,fill=gray!20, below = 7mm of ecu] (mpc) {MPC};
			\draw [-latex] (load) -- (ecu);
			
			\path (ecu.east) -- (ecu.north east) coordinate[pos=0.56] (ecu1);
			\path (ecu.east) -- (ecu.south east) coordinate[pos=0.23] (ecu2);
			
			\path (engine.west) -- (engine.north west) coordinate[pos=0.501] (eng1);
			\path (engine.west) -- (engine.south west) coordinate[pos=0.682] (eng2);
			
			\draw (ecu2) -- (inportecu);
			\draw [latex-latex] (ecu1) -- (eng1);
			\draw [-latex] (ecu) -- (mpc) node [pos = 0.6, right] {$x_k$, $y_k$};
			\draw [-latex] (outport) -- (eng2)  node [pos = 0.7, above] {$u$};

			\node [text width=5em, minimum height=3em, text centered,draw,fill=gray!20, above left = -0mm and 9mm of ecu] (modpar) {Model, tuning parameters and CT margins};
			\node [text width=5em, minimum height=3em, text centered,draw,fill=gray!20, above left = -24mm and 9mm of ecu] (maps) {Set point maps};
			\node [text width=4em,text centered, above left = -5mm and 1mm of ecu, inner sep = 0pt] {$\omega_e,$ $\dot{m}_f$};
			\draw [-latex] (ecu) -- ++(-2.5,0) -| (maps) ;
			\draw [-latex] (ecu) -- ++(-2.5,0) -| (modpar) ;
			\draw [-latex] (maps.east) -- ++(0.5,0) |- (mpc.146); 
			\draw [-latex] (modpar.west) -- ++(-0.3,0) |- (mpc.210) ;
			\node [text width=10em, minimum height=2em, above left = -22mm and -1mm of mpc, inner sep = 0pt] {$A^{g}$, $B^{g}$, $C^{g}$, $D^{g}$, $\tau_{\boost}^{g}$, $\tau_{\EGR}^{g}$, $w^{g}$, $\boldsymbol{\epsilon}^{g}$, $\boldsymbol{\sigma}^{g*}$, $\boldsymbol{\mu}^{g*}$};
			\node [text width=2em, minimum height=3em, above left = -5mm and 0mm of mpc, inner sep = 0pt] {$\steady{u}_{k}$, $\steady{x}_{k}$, $\steady{y}_{k}$};
			\path [draw] (mpc.east) -- ++(0.5,0)  |- (inportmpc) ;
			\end{tikzpicture}
			\par\end{centering}
		\protect\caption{Controller configuration.}
		\label{fig:con_config}
	\end{figure}

	\begin{figure}
		\begin{centering}
			\begin{tikzpicture}
			\node  at (0,0) (pic)
			{\includegraphics[clip, trim = {0.cm 4.75cm 0cm 0cm}]{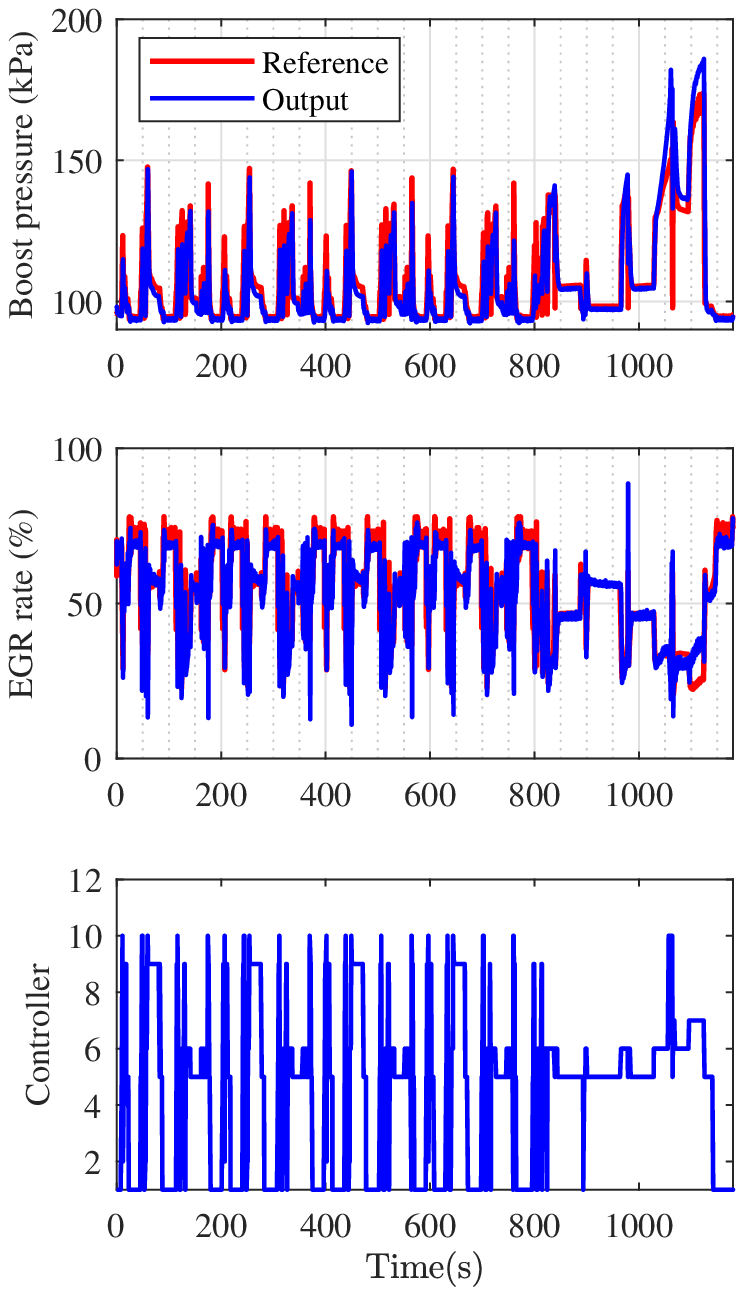}};	
			
			%% Normalising the axis %%
			
			\node (rect) [draw =none , fill = white, minimum width = 1.1cm, minimum height = 7.75cm, inner sep = 0pt, below  left = -84mm and -11.5mm of pic] {};
			
			\node [ below  left = -85mm and -12mm of pic] {\small{\pgfmathparse{200/\pimbar}\pgfmathprintnumber\pgfmathresult}};
			\node [ below  left = -70.5mm and -12mm of pic] {\small{\pgfmathparse{150/\pimbar}\pgfmathprintnumber\pgfmathresult}};
			\node [ below  left = -56.5mm and -12mm of pic] {\small{\pgfmathparse{100/\pimbar}\pgfmathprintnumber\pgfmathresult}};
			
			\node [ rotate = 90, below left = -81mm and 0.5mm of pic] {\small{Normalised p$_{\textrm{im}}$}};
			
			\node [ below  left = -42mm and -12mm of pic] {\small{\pgfmathparse{100/\egrbar}\pgfmathprintnumber\pgfmathresult}};
			\node [ below  left = -25mm and -12mm of pic] {\small{\pgfmathparse{50/\egrbar}\pgfmathprintnumber\pgfmathresult}};
			\node [ below  left = -10mm and -12mm of pic] {\small{\pgfmathparse{0/\egrbar}\pgfmathprintnumber\pgfmathresult}};
			
			\node [ rotate = 90, below left = -36mm and 0.5mm of pic] { \small{Normalised {y$_{\textrm{EGR}}$}}};
			
			%% Labeling the time axis %%
			
			\node (rect) [draw =none, fill = white, minimum width = 6.7cm, minimum height = 0.3cm, inner sep = 0pt, below  left = -49mm and -77mm of pic] {};
			
			\node (rect) [draw =none, fill = white, minimum width = 6.7cm, minimum height = 0.3cm, inner sep = 0pt, below  left = -5.5mm and -77mm of pic] {};

			\node [ below  left = -50mm and -15mm of pic] {\small{0}};
			\node [ below  left = -50mm and -27mm of pic] {\small{200}};
			\node [ below  left = -50mm and -37.5mm of pic] {\small{400}};
			\node [ below  left = -50mm and -48.mm of pic] {\small{600}};
			\node [ below  left = -50mm and -58.5mm of pic] {\small{800}};
			\node [ below  left = -50mm and -70mm of pic] {\small{1000}};
			
			\node [ below  left = -6.5mm and -15mm of pic] {\small{0}};
			\node [ below  left = -6.5mm and -27mm of pic] {\small{200}};
			\node [ below  left = -6.5mm and -37.5mm of pic] {\small{400}};
			\node [ below  left = -6.5mm and -48.mm of pic] {\small{600}};
			\node [ below  left = -6.5mm and -58.5mm of pic] {\small{800}};
			\node [ below  left = -6.5mm and -70mm of pic] {\small{1000}};

			\node [ below left  = -2mm and -52mm of pic] {Time (s)};
			
			\end{tikzpicture}
			\par\end{centering}
		\protect\caption{Boost pressure and EGR rate trajectories over NEDC.}
		\label{fig:nedc_empc_final}
	\end{figure}

	\begin{figure}
	\begin{centering}
		\begin{tikzpicture}
		\node  at (0,0) (pic)
		{\includegraphics[clip, trim = {0.cm 4.75cm 0cm 0cm}]{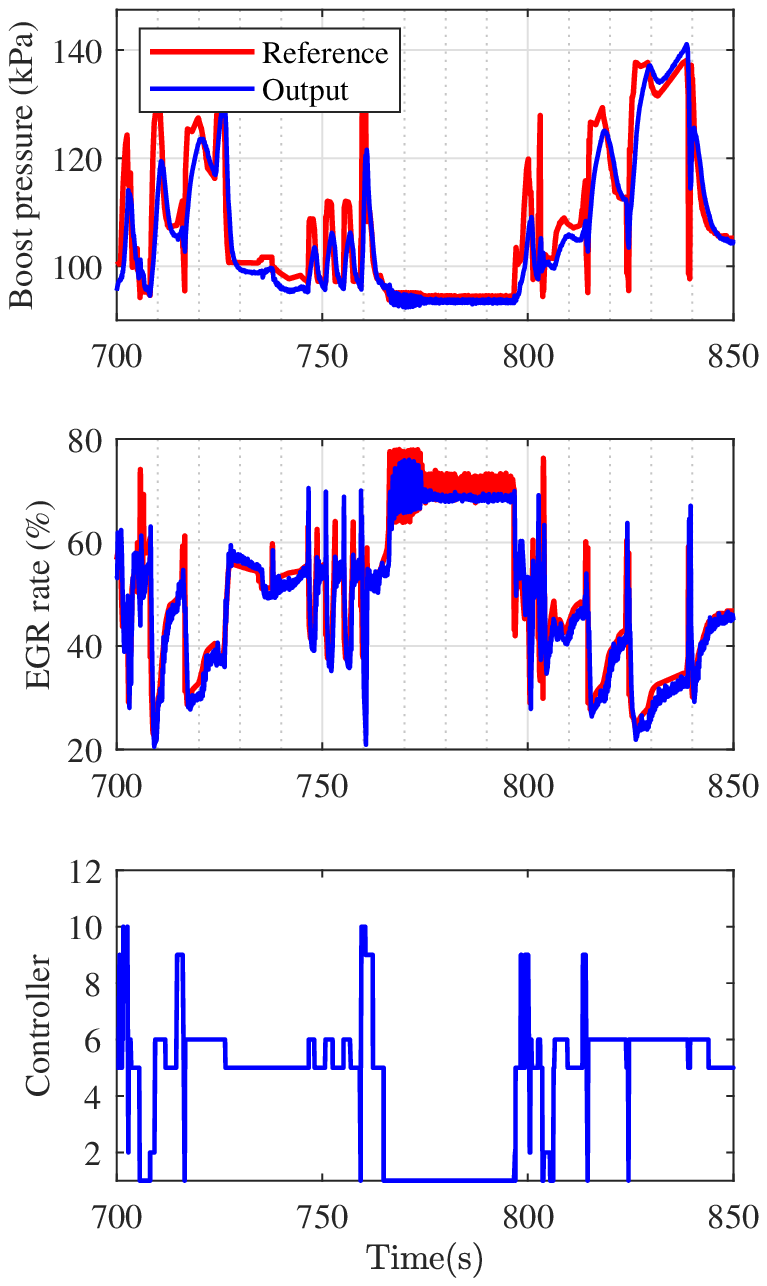}};	
		
		%% Normalising the axis %%
		
		\node (rect) [draw =none , fill = white, minimum width = 1.1cm, minimum height = 7.75cm, inner sep = 0pt, below  left = -84mm and -11.5mm of pic] {};
		
		\node [ below  left = -81mm and -12mm of pic] {\small{\pgfmathparse{140/\pimbar}\pgfmathprintnumber\pgfmathresult}};
		\node [ below  left = -70mm and -12mm of pic] {\small{\pgfmathparse{120/\pimbar}\pgfmathprintnumber\pgfmathresult}};
		\node [ below  left = -59mm and -12mm of pic] {\small{\pgfmathparse{100/\pimbar}\pgfmathprintnumber\pgfmathresult}};
		
		\node [ rotate = 90, below left = -81mm and 0.5mm of pic] {\small{Normalised p$_{\textrm{im}}$}};
		
		\node [ below  left = -42mm and -12mm of pic] {\small{\pgfmathparse{80/\egrbar}\pgfmathprintnumber\pgfmathresult}};
		\node [ below  left = -31mm and -12mm of pic] {\small{\pgfmathparse{60/\egrbar}\pgfmathprintnumber\pgfmathresult}};
		\node [ below  left = -20mm and -12mm of pic] {\small{\pgfmathparse{40/\egrbar}\pgfmathprintnumber\pgfmathresult}};
		\node [ below  left = -10mm and -12mm of pic] {\small{\pgfmathparse{20/\egrbar}\pgfmathprintnumber\pgfmathresult}};
		
		\node [ rotate = 90, below left = -36mm and 0.5mm of pic] { \small{Normalised {y$_{\textrm{EGR}}$}}};
		
		%% Labeling the time axis %%
		
		\node (rect) [draw =none, fill = white, minimum width = 6.9cm, minimum height = 0.3cm, inner sep = 0pt, below  left = -49mm and -79mm of pic] {};
		
		\node (rect) [draw =none, fill = white, minimum width = 6.9cm, minimum height = 0.3cm, inner sep = 0pt, below  left = -5.5mm and -79mm of pic] {};

		\node [ below  left = -50mm and -18mm of pic] {\small{700}};
		\node [ below  left = -50mm and -37.5mm of pic] {\small{750}};
		\node [ below  left = -50mm and -58.5mm of pic] {\small{800}};
		\node [ below  left = -50mm and -77mm of pic] {\small{850}};
		
		\node [ below  left = -6.5mm and -18mm of pic] {\small{700}};
		\node [ below  left = -6.5mm and -37.5mm of pic] {\small{750}};
		\node [ below  left = -6.5mm and -58.5mm of pic] {\small{800}};
		\node [ below  left = -6.5mm and -77mm of pic] {\small{850}};

		\node [ below left  = -2mm and -52mm of pic] {Time (s)};
		
		\end{tikzpicture}
		\par\end{centering}
	\protect\caption{Magnified view of Fig. \ref{fig:nedc_empc_final} from $\SI{700}{s}$ to $\SI{850}{s}$.}
	\label{fig:nedc_empc_final_mag}
\end{figure}

%============================================================================================%
\section{Experimental results}
\label{sec:exp_res}

\subsection{Real Time Implementation}

For experimental validation of the proposed control formulation, a test bench at Toyota's Higashi-Fuji Technical Center in Susono, Japan is used. The test bench is equipped with a diesel engine and a transient dynamometer and the controller proposed in Section \ref{sec:controller} is implemented in real-time using the dSPACE DS1006 real-time processor board \cite{DS1006}. The controller configuration is shown in Fig.~\ref{fig:con_config}. Based on the switched LTI-MPC strategy, for the current engine speed and fuelling rate,  $\left(\omega_e,\dot{m}_f\right)$, the model, tuning parameters and CT margins are selected at each time instant as shown in Fig.~\ref{fig:con_config}. The MPC optimisation problem \eqref{eq:mpc} is expressed in the condensed form and the quadratic programming in C (QPC) suite \cite{Wills2012} is chosen for solving by using the interior-point \texttt{qpip} method.

\begin{figure}
	\begin{centering}
		\begin{tikzpicture}
		\node at (0,0) (pic)
		{\includegraphics[clip, trim = {0.6cm 0cm 0cm 0cm}]{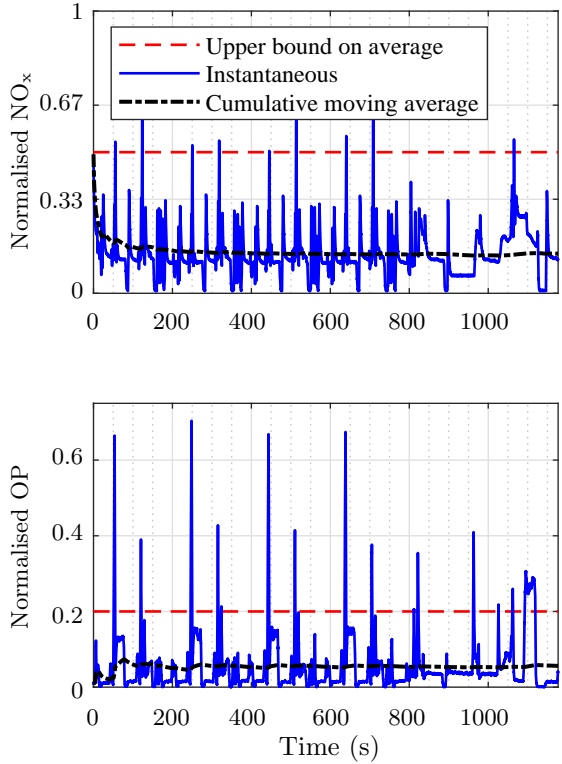}};		
		
		%% Normalising the axis %%
		
		\node (rect) [ draw = none, fill = white, minimum width = 0.9cm, minimum height = 9.5cm, inner sep = 0pt, below  left = -103mm and -7.8mm of pic] {};
		
		\node [ below  left = -103mm and -8mm of pic] {\small{\pgfmathparse{600/\NOxbar}\pgfmathprintnumber\pgfmathresult}};
		\node [ below  left = -91mm and -8mm of pic] {\small{\pgfmathparse{400/\NOxbar}\pgfmathprintnumber\pgfmathresult}};
		\node [ below  left = -78mm and -8mm of pic] {\small{\pgfmathparse{200/\NOxbar}\pgfmathprintnumber\pgfmathresult}};
		\node [ below  left = -66mm and -8mm of pic] {\small{\pgfmathparse{0/\NOxbar}\pgfmathprintnumber\pgfmathresult}};
		
		\node [ rotate = 90, below left = -94mm and 4.0mm of pic] {\small{Normalised {NO$_{\textrm{x}}$}}};
		
		\node [ below  left = -44mm and -8mm of pic] {\small{\pgfmathparse{60/\OPbar}\pgfmathprintnumber\pgfmathresult}};
		\node [ below  left = -34mm and -8mm of pic] {\small{\pgfmathparse{40/\OPbar}\pgfmathprintnumber\pgfmathresult}};
		\node [ below  left = -23mm and -8mm of pic] {\small{\pgfmathparse{20/\OPbar}\pgfmathprintnumber\pgfmathresult}};
		\node [ below  left = -13mm and -8mm of pic] {\small{\pgfmathparse{0/\OPbar}\pgfmathprintnumber\pgfmathresult}};
		
		\node [ rotate = 90, below left = -42mm and 4mm of pic] { \small{Normalised {OP}}};
		
		%% Labeling the time axis %%
		
		\node (rect) [draw = none , fill = white, minimum width = 6.3cm, minimum height = 0.3cm, inner sep = 0pt, below  left = -61.3mm and -70mm of pic] {};
		
		\node (rect) [draw = none , fill = white, minimum width = 6.3cm, minimum height = 0.8cm, inner sep = 0pt, below  left = -9.3mm and -70mm of pic] {};

		\node [ below  left = -62mm and -10mm of pic] {\small{0}};
		\node [ below  left = -62mm and -22mm of pic] {\small{200}};
		\node [ below  left = -62mm and -32mm of pic] {\small{400}};
		\node [ below  left = -62mm and -43.mm of pic] {\small{600}};
		\node [ below  left = -62mm and -53.mm of pic] {\small{800}};
		\node [ below  left = -62mm and -64.mm of pic] {\small{1000}};
		
		\node [ below  left = -9.5mm and -10mm of pic] {\small{0}};
		\node [ below  left = -9.5mm and -22mm of pic] {\small{200}};
		\node [ below  left = -9.5mm and -32.0mm of pic] {\small{400}};
		\node [ below  left = -9.5mm and -43.mm of pic] {\small{600}};
		\node [ below  left = -9.5mm and -53.0mm of pic] {\small{800}};
		\node [ below  left = -9.5mm and -64.mm of pic] {\small{1000}};

		\node [ below left  = -6mm and -47mm of pic] {Time (s)};
		
		\end{tikzpicture}
		\par\end{centering}
	\protect\caption{Instantaneous engine-out NO$_\textrm{x}$ emissions and opacity, and the cumulative moving average over NEDC.}
	\label{fig:em_nedc}
\end{figure}

\subsection{Results and discussions}	
	 In this section, the experimental results obtained by implementing the proposed MPC with transient average emissions constraints over new European driving cycle (NEDC) are presented. The baseline parameters of the controllers are chosen identical to the simulation study in the previous section. The controllers are tuned as in \cite{Sankar2018} and the closed-loop response obtained over the NEDC with the choice of final tuning parameters is shown in Fig. \ref{fig:nedc_empc_final}. Fig.~\ref{fig:nedc_empc_final_mag} shows the tracking performance in both output channels over \SI{150}{\second} of NEDC.

\subsubsection{Satisfaction of average emissions constraints}

The NO$_\textrm{x}$ level and opacity of the engine-out exhaust gas are measured using Horiba MEXA~1600D DEGR system and AVL~439 opacimeter, respectively. The instantaneous NO$_\textrm{x}$ emissions and opacity, and their cumulative moving average over the drive cycle are shown in Fig.~\ref{fig:em_nedc}. The upper bound on the NO$_\textrm{x}$ emissions and opacity averaged over the duration of the drive cycle are chosen identical to that used in the first case in simulation study in Section~\ref{sec:sim}, namely $\pgfmathparse{300/\NOxbar}\pgfmathprintnumber\pgfmathresult$ and $\pgfmathparse{20/\OPbar}\pgfmathprintnumber\pgfmathresult$, respectively. From Fig.~\ref{fig:em_nedc}, it can be noted that the instantaneous NO$_\textrm{x}$ emissions surpass $\pgfmathparse{300/\NOxbar}\pgfmathprintnumber\pgfmathresult$ at several time instants over the drive cycle, and opacity is greater than $\pgfmathparse{20/\OPbar}\pgfmathprintnumber\pgfmathresult$ for approximately $\SI{50}{s}$ during the drive cycle. However, the cumulative moving averages of NO$_\textrm{x}$ and opacity adhere to their upper bound constraints as seen in Fig.~\ref{fig:em_nedc}. The average NO$_\textrm{x}$ emissions and opacity over the NEDC are $\pgfmathparse{84.29/\NOxbar}\pgfmathprintnumber\pgfmathresult$ and $\pgfmathparse{5.62/\OPbar}\pgfmathprintnumber\pgfmathresult$, respectively.

\subsubsection{Comparison with other MPC schemes}
The performance of the proposed controller with respect to improving fuel efficiency is evaluated by comparing three different controllers.
\begin{enumerate}
	\item MPC-A: The MPC formulation used by \cite{Sankar2018}.
	\item MPC-PL: The proposed controller with penalty on the pumping loss and without constraints on the average emissions i.e., the MPC formulation as in \eqref{eq:mpc} without the constraint \eqref{eq:avg1}.
	\item MPC-EPL: The proposed controller including the pumping loss penalty and average constraints on emissions i.e., the MPC formulation as in \eqref{eq:mpc}.
\end{enumerate}

\begin{table}
	\centering
	\protect\caption{Normalised transient pumping loss obtained with different controllers.}
	\par
	\centering{}
	\begin{tabularx}{0.4\columnwidth}{@{} lX @{}}
		\hline
		Controller & NTPL \tabularnewline
		\hline
		MPC-A & 1 \tabularnewline
		MPC-PL & $0.972$  \tabularnewline
		MPC-EPL & $1.034$  \tabularnewline
		\hline
	\end{tabularx}\label{tab:rmpc_empc}
\end{table}

The normalised transient pumping loss (NTPL) over NEDC is given by
\begin{align}
NTPL = \frac{\sum_{k=0}^{T} \left(p_{\textrm{em}}\left(k\right)-p_{\textrm{im}}\left(k\right)\right)}{\sum_{k=0}^{T} \left(p_{\textrm{em}}^{\textrm{MPC-A}}\left(k\right)-p_{\textrm{im}}^{\textrm{MPC-A}}\left(k\right)\right)}\label{eq:perf_metric}
\end{align}

\noindent where $T$ is the final time of NEDC; $p_{\textrm{im}}^{\textrm{MPC-A}}\left(k\right)$ and 
$p_{\textrm{em}}^{\textrm{MPC-A}}\left(k\right)$ are the intake and exhaust manifold pressures obtained at some time instant $k$ with MPC-A over NEDC, respectively. In this work, pumping loss is used as an analogue for fuel consumption because: (i) fuelling rate is not measured and (ii) the fuelling rate estimator used in the ECU ignores changes in the fuel rail pressure and other external factors affecting the fuelling rate.

As a result of explicit penalty of pumping loss in the MPC-PL formulation, the transient pumping loss incurred is lower compared to that obtained with MPC-A. Considering the complete drive cycle, the normalised transient pumping loss (NTPL) achieved with MPC-PL is $2.8\%$ lower than that obtained with MPC-A, which translates to better fuel efficiency. On the other hand, with MPC-EPL, despite minimising the transient pumping loss, the presence of the upper bound constraint on the average emissions has resulted in an increase of pumping loss with an NTPL of $1.034$ (see Table~\ref{tab:rmpc_empc}) compared to MPC-A, i.e., more fuel is consumed by incorporating the average emissions constraints.

%============================================================================================%
\section{Conclusions}
\label{sec:conclusions}

	In this paper, a model predictive controller with average constraints on the emissions has been proposed for diesel engine airpath and experimentally demonstrated. Steady state engine-out NO$_\textrm{x}$ and opacity maps were developed as a function of states and inputs. The emissions averaged over the drive cycle were constrained to remain within certain upper bound in the MPC formulation. In addition, transient pumping loss was minimised explicitly in the controller, which is shown to improve the fuel economy. Furthermore, the controller has a reduced set of effective tuning parameters compared to the conventional MPC to aid rapid calibration. 
	
	Good reference tracking in both output channels using the proposed controller architecture has been experimentally demonstrated over NEDC. The NO$_\textrm{x}$ level and opacity of the engine-out exhaust gas averaged over NEDC have been shown to satisfy their upper bounds. Comparison of the pumping loss over NEDC using the proposed controller with and without the average emissions constraints corroborates the existence of a trade-off between fuel consumption and emissions. 
	
	Further research can develop strategies that adapts the upper bound on the averaged emissions based on estimated averaged speed over an appropriate time period. This will help in implementation of the averaged emissions constraints in real world driving.

%============================================================================================%
\section*{Acknowledgements}
The authors would like to thank the engineering staff at Toyota's Higashi-Fuji Technical Center in Susono, Japan, for assisting with the experiments.

%============================================================================================%
\appendix
\section*{Appendix}
\renewcommand{\thesubsection}{\Alph{subsection}}
\renewcommand\theequation{A.\arabic{equation}}

\subsection{Proof of Theorem \ref{thm:feas}}
\begin{proof} [\unskip\nopunct]
	Let	the optimal control sequence for \eqref{eq:mpc} and the corresponding state sequence at time $k$ be $\boldsymbol{\pv{u}}_k^*=\left\{\pv{u}_{k|k}^*,\,\ldots,\,\pv{u}_{k+N-1|k}^*\right\}$ and $\boldsymbol{\pv{x}}_k^*=\left\{\pv{x}_{k|k}^*,\,\ldots,\,\pv{x}_{k+N|k}^*\right\}$, respectively. Because of feasibility of \eqref{eq:mpc} at time $k$, the tightened constraints and the terminal constraint are satisfied i.e., $\pv{u}_{k+j|k}^* \in \mathcal{U}^g_j$,  $\pv{x}_{k+j|k}^* \in \mathcal{X}_j^g$ $\forall j\in\mathbb{Z}_{\left[0:N-1\right]}$ and $\pv{x}_{k+N|k}^* \in \mathcal{X}_f$. Consider the following candidate control sequences for \eqref{eq:mpc} at the time step $k+1$,  $\boldsymbol{\pv{u}}_{k+1}^0=\left\{\pv{u}_{k+1|k+1}^0,\,\ldots,\,\pv{u}_{k+N|k+1}^0\right\}$, where 
	\begin{subequations}
		\begin{align}
		\ltvpv{u}_{k+1+j|k+1}^0 &= \ltvpv{u}_{k+1+j|k}^* + P_{j}^{g'}w_k + \boldsymbol{m}_{k+1+j|k+1} \nonumber \\ 
		& \hspace{1.8cm} + \Delta {\steady{u}},  \forall j \in \mathbb{Z}_{\left[0:N-2\right]}, \\
		\ltvpv{u}_{k+N|k+1}^0 &= \kappa_f\left(\ltvpv{x}_{k+N|k}^0\right),
		\end{align}\label{eq:inpcan}
	\end{subequations}	
	
	\noindent at prediction step $j$,  $P_{j}^{g'}$ is the disturbance feedback policy of the constraint tightening approach corresponding to the controller at ${g'}$;  $\boldsymbol{m}_{k+1+j|k+1}$ is the input perturbation at time $k+1$ added to reject the disturbance due to controller switching with $K_x$ denoting a nilpotent candidate feedback gain; and $\forall k$, $g$ and $g'$,
	
	\begin{subequations}
		\begin{align}
		\boldsymbol{m}_{k+1+j|k+1} & =   - K_x\left(\boldsymbol{e}_{k+1}+A^{g'}\boldsymbol{n}_{k+j|k+1}\right) \label{eq:m_ltv} \\
		\boldsymbol{n}_{k+1+j|k+1} & =  \begin{cases}
		\begin{array}{ll}
		\begin{array}{l}
		0,
		\end{array} & j=0\\
		\begin{array}{l}
		\boldsymbol{e}_{k+1}+A^{g'}\boldsymbol{n}_{k+j|k+1}  \\ 
		+B^{g'}\boldsymbol{m}_{k+j|k+1}
		\end{array}, & j>0
		\end{array}\end{cases} \label{eq:n_ltv} \\		
		\boldsymbol{e}_{k+1} & =  
		\left(A^{g'}-A^{g}\right)  \ltvpv{x}^*_{k+1+j|k} \nonumber \\ 
		& +\left(B^{g'}-B^{g}\right) \ltvpv{u}^*_{k+1+j|k},  \forall j \in \mathbb{Z}_{\left[0:N-1\right]},
		\end{align}
	\end{subequations}
	\noindent $\mathcal{N}_j= \left\{\boldsymbol{n}_{k+1+j|k+1} | \eqref{eq:n_ltv}\right\} \forall j \in \mathbb{Z}_{\left[0:N-1\right]}$,  $ \mathcal{M}_j = \left\{	\boldsymbol{m}_{k+1+j|k+1} | \eqref{eq:m_ltv} \right\} \forall j \in \mathbb{Z}_{\left[0:N-2\right]}$.
	
	Let the corresponding candidate state sequence be $\boldsymbol{\pv{x}}_{k+1}^0=\left\{\pv{x}_{k+1|k+1}^0,\,\ldots,\,\pv{x}_{k+1+N|k+1}^0\right\}$, with  
	\begin{subequations}
		\begin{align}	 
		\ltvpv{x}_{k+1+j|k+1}^0 &= \ltvpv{x}_{k+1+j|k}^* + L_{j}^{g'}w_k + \boldsymbol{n}_{k+1+j|k+1} \nonumber \\ 
		&\hspace{1.5cm} + \boldsymbol{s}_{j} + \Delta {\steady{x}}, \forall j \in \mathbb{Z}_{\left[0:N-1\right]}, \label{eq:ltvcan}\\
		\ltvpv{x}_{k+1+N|k+1}^0 &= A^{g'}\ltvpv{x}_{k+N|k+1}^0 + B^{g'}\ltvpv{u}_{k+N|k+1}^0 ,
		\end{align}\label{eq:stcan}
	\end{subequations}
	
	\noindent and the difference in steady state and input values between successive time steps represented by $\Delta {\steady{x}} = |{\steady{x}_{k+1}} - {\steady{x}_k}| \in \Delta\mathcal{X}$ and  $\Delta {\steady{u}} = |{\steady{u}_{k+1}} - {\steady{u}_k}| \in \Delta\mathcal{U}$, respectively. The set $\mathcal{S}_j$ is chosen such that it satisfies $\forall$ $g$ and $g'$,
	\begin{align}
	s_j &= \sum_{i=1}^j \left[\left(A^{g'} L_{j}^{g'} + B^{g'} P_{j}^{g'}\right) - \left(A^{g} L_{j}^{g} + B^{g} P_{j}^{g} \right) \right]w_k\nonumber \\
	& \hspace{4cm}  \in \mathcal{S}_j\,  \forall k,\,w_k\in \mathcal{W}^{g}.\label{eq:w_unc}
	\end{align}
	
	The  candidate solution constructed from the optimal solution of $\mathcal{P}_{N}  \left(x(k),\,g\right)$ is utilised to show feasibility of $\mathcal{P}_{N}  \left(x(k+1),\,g'\right)$. Then by induction, feasibility of  $\mathcal{P}_{N}  \left(x(k),\,g'\right)$ implies feasibility of $\mathcal{P}_{N}  \left(x(k+j),\,g'\right)$ $\forall \,j>0$. The candidate solution has been shown to satisfy the constraints \eqref{eq:initst}-\eqref{eq:dyn2}, \eqref{eq:decay}-\eqref{eq:termcon} and \eqref{eq:inpslew1}-\eqref{eq:inpslew2} of $\mathcal{P}_{N}  \left(x(k+1),\,g'\right)$ in \cite{Sankar2018}. It is left to show satisfaction of \eqref{eq:emdyn} and \eqref{eq:avg1} at time $k+1$. Now consider the following candidate solution,	
	\begin{subequations}
		\begin{align}
		& \pv{v}_{k+1+j|k+1}^0 = \pv{v}_{k+1+j|k}^*,\,  \forall j \in \mathbb{Z}_{\left[0:N-2\right]}, \label{eq:vcan1}\\ 
		& \pv{v}_{k+N|k+1}^0 = C_v \pv{x}_{k+N|k}^* + D_v \kappa_f\left(\pv{x}_{k+N|k}^*\right). \label{eq:vcan2}
		\end{align}
	\end{subequations}
	
	The constraint \eqref{eq:emdyn} is satisfied by construction. Evaluating the constraint \eqref{eq:avg1} $\forall j \in \mathbb{Z}_{\left[0:N-2\right]}$ at time $k+1$ with the above candidate solution corresponds to constraint \eqref{eq:avg1} $\forall j \in \mathbb{Z}_{\left[1:N-1\right]}$ at time $k$. As ${x}_{k+N|k}^* \in \mathbb{X}_f$, by Assumption \ref{ass:terminal_cntrl} and \eqref{eq:vcan2}, \eqref{eq:avg1} is satisfied at $k+1$. Hence, the optimisation problem \eqref{eq:mpc} is recursively feasible.
\end{proof}

\setcounter{equation}{-1}
\renewcommand\theequation{B.\arabic{equation}}
\subsection{Proof of Theorem \ref{thm:stab}}

\begin{proof}[\unskip\nopunct]
	Let the optimal output, envelope sequences obtained at time $k$ be $\boldsymbol{\pv{y}}_k^*=\left\{\pv{y}_{k|k}^*,\,\ldots,\,\pv{y}_{k+N-1|k}^*\right\}$ and  $\boldsymbol{{Y}}_k^*=\left\{{Y}_{k|k}^*,\,\ldots,\,{Y}_{k+N-1|k}^*\right\}$, respectively. Since the trimming conditions remain constant, the perturbations in the candidate sequences in \eqref{eq:inpcan} and \eqref{eq:stcan} vanish (i.e., $\boldsymbol{m}_{k+1+j|k+1} = 0,\, \forall j \in \mathbb{Z}_{\left[0:N-2\right]}$, $\boldsymbol{n}_{k+1+j|k+1} = 0,\,\forall j \in \mathbb{Z}_{\left[0:N-1\right]}$ and $s_j=0$ in \eqref{eq:w_unc}). Additionally, with $N_{np}=1$, $L_j = 0,\, \forall j\in \mathbb{Z}_{\left[1:N\right]}$ and $P_j = 0,\, \forall j\in \mathbb{Z}_{\left[1:N-1\right]}$. Therefore, consider the following candidate output and envelope sequences for \eqref{eq:mpc} at the time step $k+1$, $\boldsymbol{\pv{y}}_{k+1}^0=\left\{\pv{y}_{k+1|k+1}^0,\,\ldots,\,\pv{y}_{k+N|k+1}^0\right\}$ and $\boldsymbol{{Y}}_{k+1}^0=\left\{{Y}_{k+1|k+1}^0,\,\ldots,\,{Y}_{k+N|k+1}^0\right\}$, respectively, where $\forall j \in \mathbb{Z}_{\left[0:N-2\right]}$,
	\begin{subequations}
		\begin{align*}
		& \pv{y}_{k+1+j|k+1}^0 = \pv{y}_{k+1+j|k}^*, \\
		& {Y}_{k+1+j|k+1}^0 = {Y}_{k+1+j|k}^*,\\
		& \pv{y}_{k+N|k+1}^0 = C^g \pv{x}_{k+N|k}^* + D^g \kappa_f\left(\pv{x}_{k+N|k}^*\right),\\
		& {Y}_{k+N|k+1}^0 = \Gamma^g {Y}_{k+N-1|k}^*.
		\end{align*}
	\end{subequations}	
	
	Also, the optimal pumping loss sequence for \eqref{eq:mpc} at time instant $k$ is $\left\{\pv{p}^{\textrm{loss*}}_{k|k},\,\ldots,\,\pv{p}^{\textrm{loss*}}_{k+N-1|k}\right\}$ and at time $k+1$, the corresponding candidate pumping loss sequence, $\left\{\pv{p}^{\textrm{loss}^0}_{k+1|k+1},\,\ldots,\,\pv{p}^{\textrm{loss}^0}_{k+N|k+1}\right\}$, can be constructed using \eqref{eq:ploss} and the candidate state sequence \eqref{eq:stcan}. 
	
	The cost function of $\mathcal{P}_N\left(x(k+1),\,g\right)$ is	
	%	\begin{subequations}
	\begin{align*}
	&V_{N}\left(x(k+1),\, \boldsymbol{\pv{u}}_{k+1}^0\right) = V_{N}\left(x(k),\, \boldsymbol{\pv{u}}_k^*\right)   \nonumber\\
	&  + \left\Vert W^{g}\, \Gamma^{g} Y_{k+N-1|k}^*\right\Vert^{2} - \left\Vert W^{g}\, Y_{k|k}^*\right\Vert^{2}    \nonumber\\
	&  +	\left\Vert\boldsymbol{\epsilon^{g}} W^{g}\left(\pv{y}_{k+N|k+1}^0-\Gamma^{g} \pv{y}_{k+N-1|k}^*\right)\right\Vert^{2}    \nonumber\\
	&-	\left\Vert\boldsymbol{\epsilon^{g}} W^{g}\left(\pv{y}_{k+1|k}^*-\Gamma^{g} \pv{y}_{k|k}^*\right)\right\Vert^{2} \nonumber \\ 
	& +  \alpha \left\Vert  \pv{p}^{\textrm{loss}^0}_{k+N|k+1}\right\Vert^{2}  - \alpha \left\Vert  \pv{p}^{\textrm{loss*}}_{k|k}  \right\Vert^{2}\nonumber \\ 
	&   + \gamma \left\Vert \kappa_f\left(\pv{x}_{k+N|k}^*\right) \right\Vert^{2}  - \gamma \left\Vert \pv{u}^*_{k|k} \right\Vert^{2}.
	\end{align*}
	%	\end{subequations}

	Therefore, the optimal cost,
	\begin{align}
	&V_{N}\left(x(k+1),\, \boldsymbol{\pv{u}}_{k+1}^*\right) \leq V_{N}\left(x(k+1),\, \boldsymbol{\pv{u}}_{k+1}^0\right) \nonumber\\
	&\leq V_{N}\left(x(k), \,\boldsymbol{\pv{u}}_k^*\right)    + \mathcal{O} \left(\left\Vert\boldsymbol{\epsilon^{g}}\right\Vert^{2} +\left\Vert \alpha \right\Vert^{2}+\left\Vert \kappa_f \right\Vert^{2} \right) \nonumber\\
	& - \alpha \left\Vert  \pv{p}^{\textrm{loss*}}_{k|k}  \right\Vert^{2} - \gamma \left\Vert \pv{u}^*_{k|k} \right\Vert^{2}. 
	\label{eq:thm_prf}
	\end{align}
	
	Eq.~\eqref{eq:stab} follows from \eqref{eq:thm_prf} by considering the definite positiveness of the optimal cost function $V_{N}\left(x\left(k\right),\,\boldsymbol{\pv{u}}_k^*\right) $ and its non- increasing evolution. 
\end{proof}

%============================================================================================%

\bibliographystyle{elsarticle-num}
\bibliography{ref}

\end{document}